%% file: main.tex
\documentclass[
    10pt,
    aps,
    prb,
    twocolumn,
    floatfix,
    superscriptaddress,
]{revtex4-2}

\usepackage[utf8]{inputenc}
\usepackage{times}

\usepackage[breaklinks, unicode]{hyperref}
\usepackage{url}
\usepackage{booktabs}
\usepackage{bm}

\usepackage{siunitx}
\usepackage[version=4]{mhchem}

\usepackage{amsfonts}
\usepackage{amssymb}
\usepackage{amsthm}
\usepackage{mathtools}
\usepackage{braket}

\usepackage{tikz-cd}
\usepackage{xcolor}
\usepackage{graphicx}

\input{commands}

\newcommand{\revise}[1]{{ #1}}
\newcolumntype{L}{>{$}l<{$}} 

\begin{document}

\setcounter{secnumdepth}{2} 
\hbadness=2000 


\title{
\texorpdfstring{
    Unified  topological characterization of electronic states in spin textures \\ from noncommutative $K$-theory
    }
    {
    Unified  topological characterization of electronic states in spin textures from noncommutative K-theory
    }
}

\author{Fabian R. Lux}
    \email{fabian.lux@yu.edu}
    \affiliation{\nyc}
    \affiliation{\mainz}
    
\author{Sumit Ghosh}
   \affiliation{\pgi} 
   
\author{Pascal Prass}
    \affiliation{\mainz}
    
\author{Emil Prodan}
    \affiliation{\nyc}
    
\author{Yuriy Mokrousov}
    \affiliation{\mainz}
    \affiliation{\pgi}

\date{\today}

\begin{abstract}
The nontrivial topology of spin systems such as skyrmions in real space can promote complex electronic states.
Here, we provide a general viewpoint at the emergence of topological spectral gaps in spin systems based on the methods of noncommutative $K$-theory. 
By realizing that the structure of the observable algebra of spin textures is determined by the algebraic properties of the noncommutative torus, we arrive at a unified understanding of topological electronic states which we predict to arise in various noncollinear setups. 
The power of our approach lies in an ability to categorize emergent topological states algebraically without referring to smooth real- or reciprocal-space quantities. 
This opens a way towards an educated design of topological phases in aperiodic, disordered, or non-smooth textures of spins and charges containing topological defects. 
\end{abstract}

\maketitle


\section{Introduction}
Noncollinear magnetism is central to many ideas in the field of spintronics and future information technology~\cite{Vedmedenko2020, Back2020}.
Recently, a diverse class of so-called multi-$\vec{q}$ magnets $-$ characterized by a phase-coherent superposition of multiple spin-spirals $-$ has received particular attention~\cite{Okubo2012, Takagi2018, Hirschberger2019, Fujishiro2019, Okumura2020}.
Experimentally established examples include one-dimensional (1D) helicoids in chiral magnets~\cite{Adams2012,Janoschek2013}, the two-dimensional skyrmion crystal of \ce{MnSi}~\cite{Neubauer2009} and the three-dimensional hedgehog lattice (HL) in \ce{MnGe}~\cite{Tanigaki2015}.
The interest in multi-$\vec{q}$ states is fueled by their connection to emergent electromagnetic fields which leads the way to rich electronic physics~\cite{ Bliokh2005, Fujita2011}
and can drive the opening of topological band gaps in the electronic spectrum~\cite{Hamamoto2015, Goebel2017, Goebel2018}.
In the adiabatic limit of smooth magnetization textures and strong coupling, this effect can be attributed to the real-space topology of the magnetization texture~\cite{Bruno2004, EverschorSitte2014}.
For example, a magnetic skyrmion will carry a quantized magnetic flux proportional to its topological charge~\cite{Nagaosa2013}. 
The mechanism by which a lattice of skyrmions can open a band gap can therefore be interpreted as the formation of Landau levels.
It is however unclear, to what extent this interpretation can be upheld as the adiabatic approximation looses validity.
E.g., the lattice constant of the HL phase in \ce{MnGe} is only of the order of only $\SI{3}{\nano\meter}$.
Furthermore, certain $3\vec{q}$ antiferromagnets possess topological band gaps, even though no topological index can be associated to the real-space texture~\cite{Ndiaye2019, Feng2020}.

In this work, we demonstrate that the emergence of topological electronic states in multi-$\vec{q}$ magnets is related to a fundamental restriction on the quantum mechanical observable algebra imposed by the magnetic texture. 
Our approach, which is based on noncommutative $K$-theory, encompasses the adiabatic limit and its semiclassical theory as a special case~\cite{Su2020}, but extends 
to arbitrary dimensions and makes sense for periodic as well as {\it aperiodic} spin arrangements on the atomic scale.
We reveal that the observable algebra of multi-$\vec{q}$ states is given by the universal $C^\ast$-algebra of the so-called noncommutative torus. 
As we show based on an effective model this has a profound effect on the emergence of a wealth of topological electronic states in multi-$\vec{q}$ textures, which can be categorized and understood in a unified manner. 
In particular, we relate the emergent topology to proper flavors of Chern numbers which do not rely on the smoothness of spin distribution in space, thus unraveling exotic higher-dimensional quantum Hall physics of noncollinear spin systems.
We believe that our results point a way towards a controlled design of topological states in various spin textures, whose character can be probed through the associated edge states and their dynamics arising in response to changes brought to textures experimentally in the laboratory.


\section{\texorpdfstring{The observable algebra of \\ multi-q magnets}{The observable algebra of multi-q magnets}}

\subsection{Tight-binding model}
We consider a class of Hamiltonians which assume the form of the following nearest-neighbor tight-binding model on a $d$-dimensional Bravais lattice, given by
\begin{align}
    H  = t \sum_{\braket{\vec{i},\vec{j}} \in \mathbb{Z}^{2d}} \ket{\vec{i}}\bra{\vec{j}}
    + \xc\sum_{\vec{i} \in \mathbb{Z}^d}  (\hatn (\vec{x}_\vec{i}) \cdot \bsigma) ~ \ket{\vec{i}}\bra{\vec{i}} ,
    \label{eq:hamiltonian}
\end{align}
where $\braket{}$ indicates the restriction to nearest-neighbor hoppings of strength $t$.
The vector field $\hatn\colon \mathbb{R}^d \to S^2$ describes the coupling of the electronic states to a magnetic texture via the exchange term proportional to $\xc$. 
To each site label $\vec{i} \in \mathbb{Z}^d$, we assign a real-space position $\vec{x}_\vec{i} = \sum_{l=1}^d \vec{i}_l \vec{a}_l \in \mathbb{R}^d$ in a Bravais lattice spanned by $\lbrace \vec{a}_i \rbrace$. 
The reciprocal lattice is defined accordingly via $ \vec{b}_i \cdot \vec{a}_j  = 2\pi \delta_{ij}$.
There is a natural action of the translation group $G = \mathbb{Z}^d$ on the $d$-dimensional lattice defined as
$
    \hat{T}_\vec{m} \ket{\vec{k}, \sigma}  =  \ket{\vec{k}+ \vec{m}, \sigma}
$.
While the hopping term is invariant under this operation, the exchange term is generally not.
The loss of translational symmetry renders the situation hopeless, as the space of possible spectra of the Hamiltonian is too large to allow for sensible classification.

\begin{figure*}[t!]
    \centering
    \includegraphics[width=\linewidth]{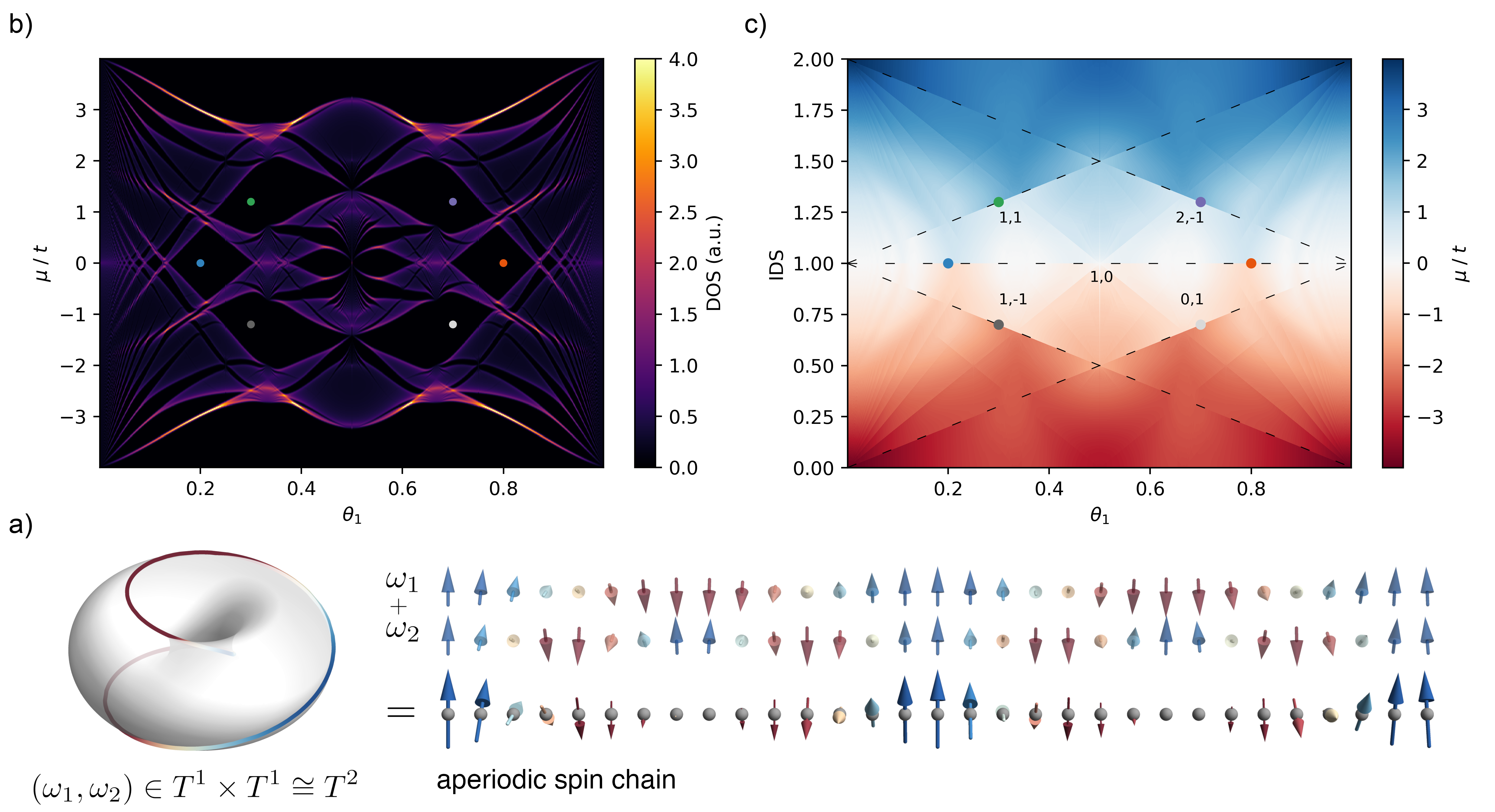}
  \caption{ 
  Fractal spectrum of a spin helix superposition. 
  Fig.~a) illustrates the superposition of two spin helix states with $\theta_1 = 2\theta_2$.
  The evolution of phase factors $(\omega_1, \omega_2)$ along the lattice generates a path on the 2-torus $T^2$.
  In Fig.~b), we calculate the DOS at $k_\mathrm{B} T = 0.01 t$ and $\xc / t =-1$ for a system with $N=1024$ sites and periodic boundary conditions, with $\theta_1=q/N$ and $q\in \mathbb{N}$, $q \leq N$.
  As $\theta_1$ increases from zero, the spectrum branches into a fractal shape reminiscent of the Hofstadter butterfly.
  Gaps open in the system, some of them are labeled by the colored bullet points. 
  Each gap can be characterized from the IDS in Fig c): discontinuities in the colormap correspond to gaps in Fig b). 
  The emerging line features are the characteristic fingerprints of the underlying $K$-theory description.
  }  
  \label{fig:limacon}
  \end{figure*}

There is an important subclass of realistic magnetic textures whose spectra are completely characterized by the topological properties of their observable algebras.
Namely, it is not uncommon that such a magnetic texture is described by one or more phase factors through which the real-space dependence will enter the Hamiltonian. 
These are generally known as multi-$\vec{q}$ states and encompass 1D textures such as spin-spirals, but also magnetic skyrmion lattices such as the famous $A$ phase of \ce{MnSi}~\cite{Neubauer2009}.
Each multi-$\vec{q}$ texture is characterized by
the presence of $r$ distinct vectors $\vec{q}_i$ with $i=1,\cdots,r$ expressed in terms of the reciprocal lattice as $\vec{q}_i = \sum_{j=1}^d \theta_{ij} \vec{b}_j$.
The dependence on the $\vec{q}$-vectors enters the magnetization texture $\hatn$ via the
phase factors 
\revise{
\begin{equation}
    \omega_i (\vec{x}_\vec{k}) \equiv  (\vec{x}_\vec{k}  \cdot \vec{q}_i/ (2\pi) + \varphi_i) \mod 1,
\end{equation}}
where $\varphi_i \in \mathbb{R}$ represents a constant phase shift, implying that instead of $\hatn(\vec{x})$  one can write $\hatn( \boldsymbol{\omega} (\vec{x}) )$.
Since $\omega_i \in [0,1) \cong \mathbb{R}/ \mathbb{Z} \cong T^1$, where $T^1$ is the 1D torus, a multi-$\vec{q}$ texture is really a composition of maps: 
\revise{\begin{equation}
    \mathbb{R}^d \overset{\boldsymbol{\omega}}{\rightarrow} T^r
     \overset{\hatn}{\rightarrow} S^2,
\end{equation}}
where $T^r$ is the $r$-dimensional torus.
By inserting the Bravais lattice expansion, one obtains
\revise{\begin{equation}
    \omega_i (\vec{x}_\vec{k}) = \left(
      \vec{k} \cdot \boldsymbol{\theta}_i
     + \varphi_i
    \right) \mod 1,
\end{equation}}
where $\boldsymbol{\theta}_i$ is the $i$-th row vector of $\theta_{ij}$.
There is also a natural action $\tau$ of the translation group $G=\mathbb{Z}^d$ on these phase factors, given by
\revise{\begin{equation}
    \tau_{\vec{m}}\omega_i (\vec{x}_\vec{k}) =\omega_i (\vec{x}_\vec{k} ) - ( \vec{m} \cdot \boldsymbol{\theta}_i \mod 1) ,
\end{equation}}
with $\vec{m}\in \mathbb{Z}^d$ and where the result is to be understood w.r.t. $\mod 1$.
This means that the pattern of phase factors is completely specified by defining the phases  $\boldsymbol{\phi} \equiv \boldsymbol{\omega}(\vec{x}_0) \in T^r$  at one arbitrary, but fixed reference point $\vec{x}_0 \in \mathbb{R}^d$.
The collection of all phase factors which are realized in the system defines {\it the hull} of the magnetic pattern~\cite{Bellissard2000}:\revise{\begin{equation}
    \Omega = G\boldsymbol{\phi}= \lbrace \tau_{\vec{m}}\boldsymbol{\phi} ~|~ \vec{m} \in \mathbb{Z}^d \rbrace  \subset T^r.
\end{equation}}
If at least one $\theta_{ij}$ is irrational, $\Omega$ forms a dense subset of $T^r$.


\subsection{The noncommutative torus}
The Hamiltonian $H$ from Eq.~(\ref{eq:hamiltonian}) can now be rewritten in a way that makes the observable algebra apparent.
The basic idea is to distill the generators of the algebra by formulating $H$ in terms of these.
This can be achieved by casting $H$ into the form 
\begin{equation}
    H =  \sum_{\vec{i} \in \mathbb{Z}^d} \hat{T}_\vec{i}  \sum_{\vec{j} \in \mathbb{Z}^d} h_\vec{i}( \boldsymbol{\phi} + \theta \vec{j} ) \ket{ \vec{j}} \bra{ \vec{j}} ,
    \label{eq:covariant_hamiltonian}
\end{equation}
where $h\colon T^r \to \mathrm{Mat}_{2\times2} (\mathbb{C}^2)$,
the sum over $\vec{j}$ visits all lattice sites, and the sum over $\vec{i}$ all neighbors.
\revise{A detailed derivation can be found in appendix \ref{app:covariant_form}.
Similar} reformulations are known from the Hofstadter model
\revise{\begin{align}
    H_\mathrm{HF} = t \sum_{\braket{\vec{i},\vec{j}} \in \mathbb{Z}^{2} \times \mathbb{Z}^2}e^{i a_{\vec{i} \vec{j}}} \ket{ \vec{i}} \bra{ \vec{j}}, 
\end{align}}
describing electrons in a uniform magnetic field~\cite{Hofstadter1976}, where $a_{\vec{i} \vec{j}}$ is the integral of the magnetic vector potential from site $\vec{x}_\vec{i}$ to site $\vec{x}_\vec{j}$.
One can rewrite
\revise{\begin{align}
    H_\mathrm{HF} = \hat{S}_1 +  \hat{S}_2 + \hat{S}_1^\dagger +  \hat{S}_2^\dagger,
\end{align}}
in terms of the magnetic translation operators $\hat{S}_i = e^{i a_{i+1,i}} \hat{T}_i$~\cite{Zak1964}, where $\hat{T}_i$ to represents a unit lattice translation in direction $i$.
These operators obey $\hat{S}_1 \hat{S}_2 = e^{2\pi i n_\Phi} \hat{S}_2 \hat{S}_1$, where $n_\Phi$ is the number of magnetic flux quanta per unit cell.
The knowledge of this algebraic structure completely characterizes the topological properties of the associated Hofstadter spectrum~\cite{Hofstadter1976, Bellissard1994, Prodan2016}.
It is known that in the limit of smooth textures and strong coupling, the more general Hamiltonian of Eq.~(\ref{eq:covariant_hamiltonian}) can be effectively mapped onto the Hofstadter Hamiltionian \cite{Hamamoto2015,Su2020}.

Eq.~(\ref{eq:covariant_hamiltonian}) demonstrates, that the observable algebra of the aperiodic spin system is completely determined (or generated) by the lattice translation operators and the space of continuous, matrix-valued functions on the torus.
In other words, any observable can be written in the same generic form as Eq.~(\ref{eq:covariant_hamiltonian}).
We can narrow down on the minimal set of generators: 
by Fourier decomposition, scalar complex functions on the torus $T^r$ are generated by the complex phase factors $u_k = e^{ 2 \pi i \phi_k}$, where $\boldsymbol{\phi} \in T^r $.
Consider $\tau_i$ to be a unit lattice translation in direction $i$. 
Following the previous definition of the translation operator, we have $ \tau_l \phi_k = \phi_k - \theta_{kl}$  and find
the commutation relations $ [\tau_i, \tau_j] = 0 $, $ [u_i, u_j]  = 0$ and $\tau_l u_k = e^{-2 \pi i \theta_{kl}} u_k  \tau_l$.
\revise{A short derivation of these commutation relations can be found in appendix \ref{app:commutation_relations}}.
By defining $\boldsymbol{\alpha}=(\tau_1, \ldots, \tau_d, u_1,\ldots u_r )$, these relations can be summarized to $\alpha_l \alpha_k = e^{2 \pi i \Theta_{lk}} \alpha_k  \alpha_l$, where
\begin{equation}
    \Theta = \begin{pmatrix}
    0 & -\theta^T \\
    \theta & 0
    \end{pmatrix} .
\end{equation}
The observables of the system can therefore be characterized by the universal $C^\ast$-algebra given by the presentation. 
\begin{align}
    \mathcal{A}_\Theta &= \Braket{ \alpha_1, \cdots, \alpha_{r+d}
    |  \alpha_l \alpha_k = e^{ 2 \pi i \Theta_{lk}} \alpha_k  \alpha_l }.
\end{align}
\revise{The defining equation uses a common shorthand notation whereby the elements before the $|$-sign are the abstract generators of the algebra and the terms following the $|$-sign are relations these generators have to fulfill.}
This algebra $\mathcal{A}_\Theta$ is known as the \emph{noncommutative torus} in $r+d$ dimensions~\cite{Rieffel1981, Connes1994, Prodan2019}.
It can be interpreted as a generalization of the algebra describing the Hofstadter model, where $\Theta_{lk}$ describes generalized magnetic fluxes in the higher-dimensional space with $r$ artificial extra dimensions associated to the $\vec{q}$-vectors~\cite{Kraus2013, Ma2021}.
\revise{
For the $C^\ast$-algebraic details we refer to \cite{Liu2022} and \cite{Prodan2016}. Appendix \ref{app:operator_algebra} gives a brisk overview for convenience.
}

\subsection{Topological classification}
$K$-theory classifies projection operators which can arise from this observable algebra~\cite{Park2008,Blackadar1998, Connes1994}.
Loosely speaking, two projections operators $P$ and $P'$ belong to the same equivalence class $[P]$ if they are related by a unitary transformation $P' = U P U^\dagger $. 
The set of all $[P]$ defines the so-called $K_0$ group of the algebra $\mathcal{A}_\Theta$.
\revise{Equivalently, one could say that $P$ and $P'$ are equivalent if there exists a continuous path between them.
We refer also \ref{app:k_theory} for a more nuanced look into the mathematics involved.} 
For the case of $\mathcal{A}_\Theta$, the $K$-theoretical properties are well-understood~\cite{Prodan2016}.
In particular, one has $K_0( \mathcal{A}_\Theta) = \mathbb{Z}^{2^{r+d-1}}$, which is a compact way of saying that any class of topologically equivalent projection operators $[P]$ can be written as a linear combination of generators $[P] =\sum_J  n_J [E_J] $, \revise{labelled by even-cardinality subsets $J$ of $\mathcal{I}=\lbrace \tau_1 \cdots, \tau_d, u_1  \cdots, u_r \rbrace$ (there are precisely $2^{r+d-1}$ such subsets).
\revise{
Loosely speaking, the coefficients $n_J \in \ZM$ count how often the generator $E_J$ occurs in $P$ up to unitary equivalence.
Once all $n_J$ are known, the class of $P$ in $K_0(\mathcal{A}_\Theta)$ is completely determined.
}

\revise{An explicit knowledge of the projections $E_J$ is not necessary to compute all $n_J$.
Rather, }
this can be done via the noncommutative $m$-th Chern numbers ~\cite{Prodan2013b, Liu2022}.
If $J'\subseteq \mathcal{I}$ with $|J'|/2 = m$, then 
\begin{equation}
    {\rm Ch}_{J'}(g) = \frac{(2\pi i)^{|J'|/2}}{( |J'|/2 )!} \sum_{\sigma \in \Ss_{|J'|}} (-1)^\sigma \Tt\Big ( P \prod_{j \in J'} \partial_{\sigma_j}P \Big ) ,
    \label{eq:chern_formula}
\end{equation}
is an $m$-th Chern number of the gap $g$ with spectral projection $P$.
$\Tt$ is the trace per unit volume and $\Ss_{|J'|}$ is the symmetric group of order $|J|$. 
What appears suggestively in the notation of derivatives $\partial_{\sigma_j}$ are so-called derivations on the observable algebra (which is the proper algebraic generalization of the concept).
The way to construct these derivations is by first observing that the commutation relations of $\mathcal{A}_\Theta$ are left invariant with respect to $\alpha_j \to \lambda_j \alpha_j$ where $\lambda_i \in \mathbb{C}$ with $|\lambda_i| = 1 $.
If we expand $A \in \mathcal{A}_\Theta$ in terms of the algebra generators, $A =  \sum_{\vec{q} \in \ZM^{r+d}} a_{\vec{q}}
\, \alpha_1^{q_1} \ldots\alpha_{r+d}^{q_{r+d}}$, we can then construct a map
\begin{equation}
    \rho_{\boldsymbol{\lambda}}(A) = \sum_{\vec{q} \in \ZM^{r+d}} a_{\vec{q}}
	\, 
	\lambda_1^{q_1} \ldots\lambda_{r+d}^{q_{r+d}}
	\alpha_1^{q_1} \ldots\alpha_{r+d}^{q_{r+d}} \in  \mathcal{A}_\Theta .
\end{equation}
This can be used to finally define $\partial_i A = \left. i \rho_{\boldsymbol{\lambda}} (A) \right|_{\boldsymbol{\lambda} \to 1}$ (see also appendix \ref{app:differential_calculus}).
The best way to gain an intuition into this technical construction is to observe what happens 
in conventional lattice periodic systems.
The Bloch functions can be generated from plane waves $e^{i \vec{k}\cdot \vec{x}}$ and the operation which sends $e^{i \vec{k}\cdot \vec{x}} \to i \vec{x} e^{i \vec{k}\cdot \vec{x}}$
is nothing but the $\vec{k}$-derivative.
One can therefore think of $\partial_{\tau_i}$ as a generalization of $\partial_{k_i}$ to the noncommutative setting.
}

Since we have found that the underlying observable algebra is the noncommutative torus, \revise{the possible Chern numbers have many interrelations and fulfill}~\cite{Prodan2016}
\begin{equation}
   {\rm Ch}_{J'} (g) =  \sum_{J \subseteq \mathcal{I} }^{|J|~\mathrm{even}}  n_J(g) \, {\rm Ch}_{J'}(E_J),
   \label{eq:chern_expansion}
\end{equation}
where ${\rm Ch}_{J'}(E_J)=1$ if $J=J'$,  ${\rm Ch}_{J'}(E_J)=0$ if $J \not\subseteq J'$ and ${\rm Ch}_{J'}(E_J) = \mathrm{Pf}( \Theta_{J\setminus J'})$ otherwise.
\revise{The operation $\mathrm{Pf}$ denotes the Pfaffian and $\Theta_{J\setminus J'}$ is the representation of $\Theta$ in the reduced index set $J\setminus J'$.
The integer coefficients $n_J(g)$ are exactly those that correspond to the decomposition of the gap projection into $K$-theory generators, i.e., $[P(g)] = \sum_J n_J(g) [E_J]$. 
If all possible ${\rm Ch}_{J'}(g)$ would be computed via Eq.~(\ref{eq:chern_formula}), this would determine all integer invariants $n_J(g)$ through Eq.~(\ref{eq:chern_expansion}).
We want to single out two relevant special cases of the expressions above that the reader might be familiar with and which have a clear physical interpretation.
The first one is the case of empty subset $J'=\emptyset$, for which
\begin{equation}
    {\rm Ch}_{\emptyset} (g) = \Tt(P)
\end{equation}
represents the integrated density of states (IDS).
The second important case corresponds to $J'=\lbrace \tau_i , \tau_j \rbrace$, where
\begin{equation}
    {\rm Ch}_{\lbrace \tau_i , \tau_j \rbrace} (g) = (2\pi i) \Tt(P [ \partial_{\tau_i} P.\partial_{\tau_j}P]).
\end{equation}
Based on the analogy of $\partial_{\tau_j}$ with the momentum space derivative in the lattice periodic setting, one could guess that this Chern number corresponds to the anomalous Hall conductivity $\sigma_{ij}$ of a band insulator in units of $e^2/h$. This is in fact the case \cite{Bellissard1994}.
}
When a periodic supercell can be chosen and the magnetic texture is smooth, all the higher-order Chern numbers can be obtained through integrals over Berry curvature expressions~\cite{Qi2008, Kraus2013, Su2020}.
\revise{In appendix \ref{app:berry_curvature}, we show how exactly the Berry curvature approach is contained in our more general framework.}

\begin{figure*}[t]
 \centering
 \includegraphics[width=\linewidth]{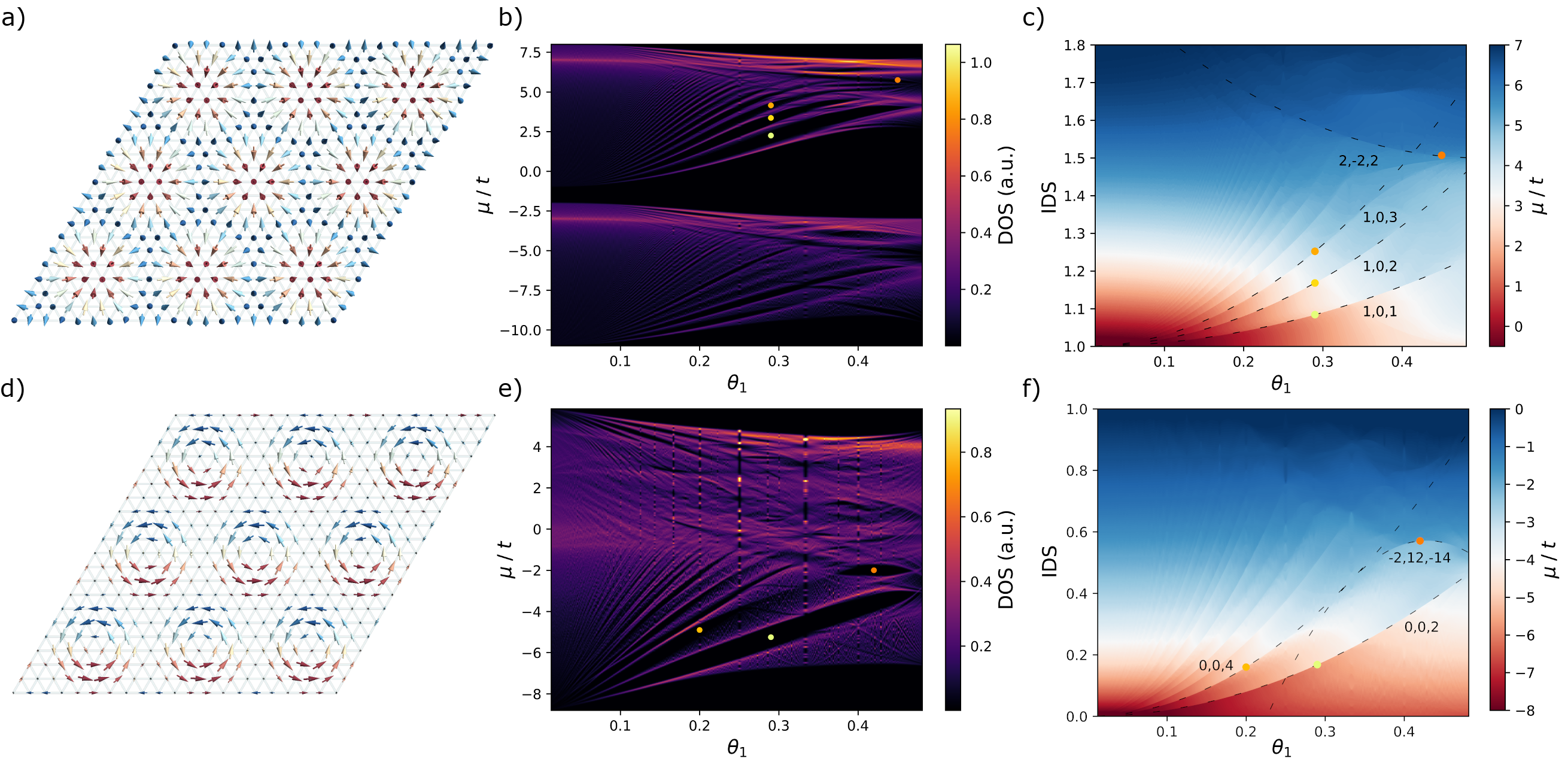}
 \caption{
     Topological spectrum of 3-$\vec{q}$-states.
     Fig a) shows an example for skyrmionic 3-$\vec{q}$-states on the triangular lattice with $N=400$, $\theta_1 = 3/\sqrt{N}$ and periodic boundary conditions.
     The DOS at $k_\mathrm{B}T = 0.01t$, $\xc / t=-5$ is calculated by combining all system sizes $\sqrt{N} \in [19,79]\cap \mathbb{N}$ with  $\sqrt{N}\theta_1 \in \mathbb{Z}$.
     A series of gaps opens in the spectrum (some of them labelled with bullet points) whose topological character is revealed in the IDS plot of Fig c).
     This procedure is repeated for the 3-$\vec{q}$ spin vortex crystal in Fig d).
    Again, the DOS in e) reveals gaps of topological character, as confirmed in f).
    The visible vertical features in e) are due to a reduced sampling density in $\theta_1$ at those points. 
 }  
 \label{fig:three_q_states}
\end{figure*}

\section{Results}
\subsection{Superposition of spin spirals}

The simplest example \revise{of all possible Chern numbers}, ${\rm Ch}_{\emptyset} (g)$, represents the integrated density of states (IDS) and we can use it to map out some possible Chern numbers that are realized by the Hamiltonian in Eq.~(\ref{eq:hamiltonian})
To illustrate this point, we start with investigating the following superposition of two helicoidal spin spirals ($r=2$):
$\hatn( \boldsymbol{\omega}(x_k ) )
=\sum_{i=1}^{2}( \cos ( 2 \pi \omega_i(x_k) )\vec{e}_y
+\sin ( 2 \pi \omega_i(x_k) )\vec{e}_z)
$,
defined on a 1D lattice ($d=1$) implying $\omega_i(x_k) = k \theta_i + \varphi_i$. 
This system could be realized physically via the proximity coupling of a 1D electronic system to two independent atomic spin spirals, each stabilizing a spin helix \revise{(one could also view it as a noncollinear spin density wave)}.
We consider the special case of $\theta_2 = 2 \theta_1$, i.e., $\theta_1$ and $\theta_2$ are rationally dependent. \revise{Other relations between $\theta_1$ and $\theta_2$ could be considered, but will not change the qualitative results, except for $\theta_1=\theta_2$, where the spectrum is trivial}.
For this situation, the orbit of the lattice translation group generates a dense, 1D subspace on the 2-torus:  $\Omega \cong T^1 \subset T^2$ which is the hull (see Fig. \ref{fig:limacon}).
For the numerical analysis, we consider a finite system of $N=1024$ atoms with periodic boundary conditions. 
The spectrum $\lbrace \epsilon_n \rbrace$ is then calculated by exact diagonalization for different values of $\theta_1$, sampled at rational values $q/N$ with $q\in \mathbb{N}$ and $q \leq N$.
In a first step, we consider the density of states at the chemical potential $\mu$, given by
$\mathrm{DOS}(\mu) = \sum_{n=1}^{2N} \partial_\mu f( \epsilon_n - \mu)$, where $f(\epsilon) = 1/ ( \exp\lbrace \epsilon / (k_B T) \rbrace + 1 ) $ is the Fermi-Dirac distribution.
The result is shown in Fig. \ref{fig:limacon} for different values of $\mu$, and reveals a characteristic Hofstadter butterfly.
As the value of $\theta_1$ is varied, multiple gaps open and close in the spectrum, some of which we label by the colored bullet points.
The topological nature of these gaps can be investigated by the means of $K$-theory.
Since the $\theta$-matrix is simply given by $\theta = (\theta_1, 2\theta_1)^T$, the evaluation of the Pfaffian leads to the prediction
\begin{equation}
    \mathrm{IDS}(g) = n_\emptyset(g) + n_{\tau u}(g)~\theta_1 ,
    \label{eq:k_prediction}
\end{equation}
with $n_{\tau u}(g) =  n_{ \lbrace \tau_1, u_1 \rbrace}(g) + 2 n_{ \lbrace \tau_1, u_2 \rbrace}(g) $, where the coefficients can be identified with the first Chern numbers 
$ \mathrm{Ch}_{ \lbrace \tau_i, u_i \rbrace} = n_{ \lbrace \tau_i, u_i \rbrace}$. 
This result can be verified by plotting the IDS versus $\theta_1$, color-coded by $\mu$ in Fig.~1.
Since the IDS is constant within a gap, the color exhibits discontinuous jumps which makes it possible to track the evolution of IDS for a specific gap as a function of $\theta_1$.
The result is in perfect agreement with Eq.~(\ref{eq:k_prediction}), and makes it possible to assign a unique label $(n_\emptyset, n_{\tau u})$ to each gap $g$.
Since the gaps with $n_{\tau u} \neq 0$ are characterized by the first Chern number, the superposition of spin helices leads to a topologically nontrivial spectrum.


\subsection{Skyrmion and vortex crystals}
One way to interpret the $\Theta$-matrix is to think of it as a generalized magnetic flux.
This is similar to the emergent field of smooth magnetic skyrmions~\cite{Bliokh2005, Schulz2012}, but is a far-reaching generalization of this concept, which also makes sense for discrete systems.
To investigate the transition into the conventional emergent field picture, we consider a triangular lattice with $\vec{a}_1 = a (1,0)^T$, $\vec{a}_2 = a (1/2,\sqrt{3}/2 )^T$ and reciprocal lattice vectors
$\vec{b}_1 = 2 \pi (1, -1/\sqrt{3})^T / a$, $	\vec{b}_2 = 2 \pi (0, 2 / \sqrt{3})^T / a$.
With respect to this lattice, we devise a $\theta$-matrix
$ \theta = \theta_1 (( 
	0, 1),
	(  1,0 ),
	( -1,-1 )) $,
which corresponds to the coherent superposition of three spin spirals. 
\revise{Technical aspects of the construction of this magnetic texture can be found in appendix \ref{app:theta_matrix_triangular_lattice}.}
This gives rise to a 3-$\vec{q}$ skyrmion lattice shown in Fig. \ref{fig:three_q_states}a) with the $K$-theory prediction
\begin{equation}
     \mathrm{IDS}(g) = n_\emptyset(g) + n_{\tau u}(g)~\theta_1 + n_{\tau^2u^2}(g)~\theta_1^2,
    \label{eq:k_prediction_skx}
\end{equation}
where $n_{\tau u} = n_{ \lbrace \tau_1 , u_2\rbrace} + n_{\lbrace\tau_1 , u_3\rbrace}+n_{\lbrace\tau_2 , u_1\rbrace}+n_{\lbrace\tau_2 , u_3\rbrace}$ and $n_{\tau^2u^2} = n_{\lbrace\tau_1, \tau_2 , u_1 , u_2\rbrace} +  n_{\lbrace\tau_1 , \tau_2 ,  u_1 ,  u_3\rbrace} + n_{\lbrace\tau_1 , \tau_2 ,  u_2  , u_3 \rbrace}$.
We proceed similarly as for the spin helices and calculate $\mathrm{DOS}(\theta_1)$ and $\mathrm{IDS}(\theta_1)$ for different system sizes.
A series of topological gaps appear in Fig.~\ref{fig:three_q_states}b) as $\theta_1$ is increased away from $0$, which can be classified by the label $(n_\emptyset, n_{\tau u}, n_{\tau^2u^2})$.
The $\theta_1^2$ dependence of the gap sizes has been theoretically observed for the skyrmion lattice~\cite{Hamamoto2015} and now finds its explanation as a fingerprint of the underlying $K$-theory.
According to the classification of Fig.~\ref{fig:three_q_states}c), 
these gaps correspond to a quantization of $n_{\tau^2u^2}$ with $n_{\tau u}=0$.
Naively, this seems to indicate that the first Chern numbers are zero which would be in contradiction to the known quantized topological Hall effect in this system.
However, the anomalous Hall effect picks up the different Chern number $\mathrm{Ch}_{ \lbrace \tau_1 \tau_2 \rbrace } = n_{ \lbrace \tau_1 \tau_2 \rbrace }$ \cite{Prodan2017} which is not visible in the IDS.
In the adiabatic limit $\xc/t \to \infty$ and $\theta_1 \to 0$, it can be shown that $\mathrm{Ch}_{ \lbrace \tau_1 \tau_2 \rbrace } = n_{\tau^2u^2}$~(see appendix \ref{app:emergent_magnetic_fields}), and the apparent contradiction is resolved.
Noteworthy, these results persist as $\theta_1$ approaches the scale of the underlying lattice, since the respective gaps can be continuously connected to the limit of $\theta_1 \to 0$.
Here (and in particular for gaps which are not connected to the adiabatic limit), any arguments based on the smoothness of the texture fail, while the $K$-theory description can still be upheld.

To further underline this point, we perform the same calculation after replacing the superposition of spin spirals by a superposition of spin density waves which gives rise to the spin texture in Fig.~\ref{fig:three_q_states}d). 
It is not obvious what the emergent field language would predict, and  
in fact, the emergent magnetic field in the continuous limit  $B_\mathrm{em} = \hatn \cdot(\partial_x \hatn \times \partial_y \hatn)$ is zero for this texture.
With the $K$-theory description, no such conceptual problems arise.
The $\theta$-matrix is not changed as compared to the previous case and the prediction for the IDS as the marker of topologically nontrivial electronic states in this system is valid. 
It is thus of no surprise that topological gaps are visible in the DOS in Fig.\ref{fig:three_q_states}d), while the associated IDS in Fig.~\ref{fig:three_q_states}e) is again in perfect agreement with the $K$-theory of multi-$\vec{q}$ order.
This serves to show that the generalized fluxes $\Theta$ can provide a connecting theme which extends to more exotic magnetic phases such as vortex crystals in frustrated magnets~\cite{Wang2015}.

\revise{
\subsection{The cubic hedgehog lattice}

Lastly, we would like to show which predictions $K$-theory could make for three-dimensional textures such as the cubic hedgehog lattice found in $\ce{MnGe}$ \cite{Tanigaki2015}.
In this case, one deals with three linearly independent, mutually orthogonal $\vec{q}$-vectors: $\vec{q}_i = q\vec{e}_i$.
This means that the $\theta$-matrix is just the identity matrix in $d=3$ dimensions: $\theta = \theta_1 \id_3$. 
A detailed list with all Chern number relations for the cubic hedgehog lattice can be found in table \ref{tab:3q_3d} of the appendix.
Here, $K$-theory tells us that the IDS of a gap $g$ should behave as a third-degree polynomial in $\theta_1$, with the highest order term controlled by third Chern number, i.e.,
\begin{equation}
	\mathrm{IDS}(g)
	=
	n_\emptyset
	+ n_{\tau u} \theta_1 
	+ n_{\tau^2 u^2} \theta_1^2
	+ n_{\tau^3 u^3} \theta_1^3,
\end{equation}
with the shorthand notation 
\begin{align}
n_{\tau u}  
& = n_{\lbrace \tau_1 u_1 \rbrace}  + n_{\lbrace \tau_2 u_2 \rbrace} + n_{\lbrace \tau_3 u_3 \rbrace} 
\\
n_{\tau^2 u^2}  
& = n_{\lbrace \tau_1 \tau_2 u_1 u_2 \rbrace}  
+ n_{\lbrace \tau_1 \tau_3 u_1 u_3 \rbrace}   
+ n_{\lbrace \tau_2 \tau_3 u_2 u_3 \rbrace}   
\\
n_{\tau^3 u^3}  
& = n_{\lbrace \tau_1 \tau_2 \tau_3 u_1 u_2 u_3 \rbrace}.
\end{align}
The latter can be identified with the top-level Chern number
\begin{equation}
	\mathrm{Ch}_{\lbrace \tau_1 \tau_2 \tau_3 u_1 u_2 u_3 \rbrace}
	=
	n_{\lbrace \tau_1 \tau_2 \tau_3 u_1 u_2 u_3 \rbrace}.
\end{equation} 
Finding a model which realizes this high-dimensional Chern number is left for future research.
}

\section{Conclusion}
In summary, we put forward a new method to characterize electronic topological states emerging in real-space spin textures based on the $K$-theory of $C^\ast$-algebras.
In contrast to conventional methods of topological characterization based on smooth Berry phase properties $-$ whose meaning is lost in aperiodic, disordered or non-smooth textures $-$ the $K$-theory analysis can be used to predict and understand the appearance of nontrivial gaps beyond this limitation. 
As such, the $K$-theory categorization bears great promise for unraveling and shaping the hybrid topological properties of complex spin textures in real materials.
Particular exciting aspects to address in the future are the topology of three-dimensional textures which have the potential to harvest six-dimensional physics (\revise{as we have shown,} nontrivial third Chern numbers are a theoretical possibility for the 3$\vec{q}$ cubic hedgehog lattice), as well as the $K$-theory topological interpretation of spin fluctuations, dynamical excitations of real-space spin systems and the associated edge state physics.
\revise{It would be further interesting to trace the evolution of topological invariants across topological magnetic phase transitions in real-space and to determine the physical observables capable of detecting this change in electronic topology.}

\acknowledgments{
We  acknowledge  funding  under SPP 2137 ``Skyrmionics'' (project  MO  1731/7-1)  of  Deutsche  Forschungsgemeinschaft (DFG) and also gratefully thank the J\"ulich Supercomputing Center and RWTH Aachen University for providing computational resources under project jiff40. 
This work was further supported by the Max Planck Graduate Center with the Johannes Gutenberg-Universit\"at Mainz (MPGC).
Emil Prodan acknowledges financial support from the W.M. Keck Foundation and USA National Science Foundation through grant  DMR-1823800.
The open source code used to generate the results of this work is available from\footnote{\url{https://github.com/luxfabian/noncommutative_torus_in_spin_systems}}.
}

\hbadness=99999 
\bibliographystyle{apsrev4-2}
%

\appendix

\begin{table*}[t!]
	\centering
	\caption{{\bf Summary of the notation in the manuscript.} 
	The following table lists important symbols and notation which is used throughout the manuscript and gives a brief explanation.
	}
	\label{tab:notation}\vspace{0.2cm}
	\begin{tabular}{c|p{15cm}}
	 Symbol  & Explanation \\ \toprule
	$t$      & Hopping parameter \\
	$\xc$    & Exchange-correlation energy \\
	$d$      & Dimensionality of the lattice \\
	$N$      & Number of sites in the lattice; Thermodynamic limit: $N\to\infty$ \\
	$\vec{a}_i$ & Bravais lattice vectors \\
	$\vec{b}_j$ & Reciprocal lattice vectors; $\vec{a}_i \cdot \vec{b}_j = 2\pi \delta_{ij}$ \\
	$\vec{x}_{\vec{k}}$ &  Real-space coordinates; $\vec{x}_{\vec{k}} = \sum_{i=1}^3 k_i \vec{a}_i $ \\
	$\vec{q}_i$ & The $i$-th $\vec{q}$-vector of the multi-$\vec{q}$ texture; 
	$\vec{q}_i = \sum_{j=1}^d \theta_{ij} \vec{b}_j$ \\
	$ \boldsymbol{\theta}_i$ & $i$-th column vector in $\theta_{ij}$;  $(\boldsymbol{\theta}_i)_j = \theta_{ij}$ \\
	$ r $  & Number of distinct $\vec{q}$-vectors \\
	$ T^r $ & The $r$-dimensional torus \\ 
	$ \omega_i (\vec{x}_\vec{k}) $ & Phase with values in $T^1$, associated with $\vec{q}_i$; $ \omega_i (\vec{x}_\vec{k}) =  (\vec{x}_\vec{k}  \cdot \vec{q}_i/ (2\pi) + \varphi_i) \mod 1 $ \\
	$\boldsymbol{\omega}$ & The collection of all $\omega_i \in T^1$ into a vector $\boldsymbol{\omega}\in T^r$ \\ 
	$\ket{\vec{k}} $ & Position ket corresponding to the atomic site at location $\vec{x}_\vec{k}$ \\
	$ \ket{\sigma} $ & Spin ket labelled by the eigenstates of Pauli matrix $\sigma_z$ \\
	$ \ket{\vec{k}, \sigma} $ & Tensor product state $ \ket{\vec{k}, \sigma} = \ket{\vec{k}} \otimes \ket{\sigma}$ \\
	$\hat{T}_{\vec{m}}$ & Lattice translation operator acting on kets with $\vec{m}\in \mathbb{Z}^d$; $  \hat{T}_\vec{m} \ket{\vec{k}, \sigma}  =  \ket{\vec{k}+ \vec{m}, \sigma}$ \\
	$\hat{T}_{i}$ & Unit lattice translation in the $i$-th direction\\
	$\tau_{\vec{m}}$ & Action of the translation group $\mathbb{Z}^d \ni \vec{m} $ on the phases;  $  \tau_{\vec{m}}\omega_i (\vec{x}_\vec{k}) = ( \omega_i (\vec{x}_\vec{k} ) - ( \vec{m} \cdot \boldsymbol{\theta}_i \mod 1) ) \mod 1 $ \\
	$\tau_{i}$ & Unit lattice translation of the phases in the $i$-th direction\\
	$\boldsymbol{\phi} $ & The phase vector at one arbitrary, but fixed reference point $\vec{x}_0 \in \mathbb{R}^d$; $\boldsymbol{\phi} \equiv \boldsymbol{\omega}(\vec{x}_0) \in T^r$ \\
	$\Omega $ & Hull of the magnetic pattern;  $\Omega =  \lbrace \tau_{\vec{m}}\boldsymbol{\phi} ~|~ \vec{m} \in \mathbb{Z}^d \rbrace  \subset T^r $  \\
	$u_k$ & Fourier amplitude $u_k = e^{ 2 \pi i \phi_k}$; generator of periodic functions (functions on the $T^r$) \\
	$\Theta $ & Generalized flux matrix; $\Theta = ( (0, -\theta^T), (\theta,0))$ \\
	$\boldsymbol{\alpha}$ & Vector of generators; $\boldsymbol{\alpha}=(\tau_1, \ldots, \tau_d, u_1,\ldots u_r )$ \\
	$\mathcal{A}_\Theta $ & The universal $C^\ast$-algebra of the noncommutative torus; $\mathcal{A}_\Theta = \Braket{ \alpha_1, \cdots, \alpha_{r+d}
    |  \alpha_l \alpha_k = e^{ 2 \pi i \Theta_{lk}} \alpha_k  \alpha_l }$\\
	$d_{\mathrm{eff}} $ & Effective dimension; $d_{\mathrm{eff}} = r+d $ \\
	$P$ & Projection operator in $\mathcal{A}_\Theta$; $P^2 = P$ \\
	$[P]$ & Equivalence class of unitarily equivalent projection operators in $\mathcal{A}_\Theta$; \\
	$K_0(\mathcal{A}_\Theta) $ & The (Grothendieck) group of all $[P]$; $K_0(\mathcal{A}_\Theta)  = \mathbb{Z}^{2^{d_\mathrm{eff}-1}}$ \\
	$\mathcal{I}$ & Index set which labels the generators; $\mathcal{I}=\lbrace \tau_1 \cdots, \tau_d, u_1  \cdots, u_r \rbrace$ 
	\\
	$J$ & An even cardinality subset of $\mathcal{I}$; $ J \subseteq \mathcal{I} \colon |J|~\mathrm{even}$ \\
	$E_J$ & Generators of $K_0(\mathcal{A}_\Theta) $ \\ 
	$\Theta_{J\setminus J'}$ & Restriction of the flux matrix to the submatrix for the restricted index set $J \setminus J'$ \\
	$\mathrm{Ch}_J(g) $ &  $|J|/2$-th Chern number of the gap with label $g$ and w.r.t. to the indices $J$ \\
	$\mathrm{Pf}(A)$ &  The Pfaffian of a matrix $A$; For $A^T = -A\colon (\mathrm{Pf}(A))^2 = \det A$ \\
	$\mathrm{IDS}(g)$ & Integrated density of states within the gap $g$; $\mathrm{IDS}(g) = \lim\limits_{N\to\infty}  \frac{1}{N} \tr~P_{E<E_g}$, where $E_g$ is an energy in $g$ and $P_{E<E_g}$ projects onto states below $E_g$
\\ \bottomrule
	\end{tabular}\\
\end{table*}

\section{Bringing the Hamiltonian into its covariant form}
\label{app:covariant_form}

We demonstrate how the Hamiltonian can indeed be written in the canonical form presented in the manuscript.
For the hopping term, one finds
\begin{align}
H_t &= t \sum_{\braket{\vec{k},\vec{l}} \in \mathbb{Z}^{2d}}\ket{\vec{k}}\bra{\vec{l}}
\notag \\
&= 
	t \sum_{\vec{k} \in \mathbb{Z}^{d}}  \sum_{l=1}^d
	\left( \ket{\vec{k}}\bra{\vec{k}+\vec{e}_l } + \ket{\vec{k}+\vec{e}_l}\bra{\vec{k} } \right)
\notag \\
&= 
	t \sum_{\vec{k} \in \mathbb{Z}^{d}} \sum_{l=1}^d
	(\hat{T}_l + \hat{T}_l^\dagger)\ket{\vec{k}}\bra{\vec{k} } 
\notag \\
&= 
 \sum_{l=1}^d
(\hat{T}_l+ \hat{T}_l^\dagger) \sum_{\vec{k} \in \mathbb{Z}^{d}} t \ket{\vec{k}}\bra{\vec{k} } 
\notag \\
&= t
\sum_{l=1}^d
(\hat{T}_l + \hat{T}_l^\dagger)  ,
\end{align}
where $\hat{T}_l$ is a unit-translation in the direction $\vec{e}_l \in \mathbb{Z}^d$.
It is therefore invariant under translations:  $\hat{T}_\vec{m} 
H_t
\hat{T}_\vec{m}^\dagger = H_t$. 
The exchange term is given by
\begin{align}
	H_\mathrm{xc} &=
	\xc\sum_{\vec{k} \in \mathbb{Z}^d}  (\hatn (\boldsymbol{\omega}(\vec{x}_\vec{k})) \cdot \bsigma) ~ \ket{\vec{k}}\bra{\vec{k}} .
\end{align}
It is not invariant under lattice translations, but transforms as
\begin{align}
	\hat{T}_\vec{m}
	H_\mathrm{xc}
	\hat{T}_\vec{m}^\dagger
	& = \xc\sum_{\vec{k} \in \mathbb{Z}^d}  (\hatn (\boldsymbol{\omega}(\vec{x}_\vec{k})) \cdot \bsigma) ~ \ket{\vec{k}+\vec{m}}\bra{\vec{k}+\vec{m}}
		\notag \\
	& = \xc\sum_{\vec{k} \in \mathbb{Z}^d}  (\hatn (\boldsymbol{\omega}(\vec{x}_{\vec{k}-\vec{m}})) \cdot \bsigma) ~ \ket{\vec{k}}\bra{\vec{k}}
	\notag \\
	& = \xc\sum_{\vec{k} \in \mathbb{Z}^d}  (\hatn (\tau_\vec{m} \boldsymbol{\omega}(\vec{x}_{\vec{k}})) \cdot \bsigma) ~ \ket{\vec{k}}\bra{\vec{k}} . 
\end{align} 
With the definition $\boldsymbol{\phi} = \boldsymbol{\omega}(\vec{x}_0)$, the exchange term can therefore also be written as 
\begin{align}
	H_\mathrm{xc}( \boldsymbol{\phi} ) &=
	\xc\sum_{\vec{k} \in \mathbb{Z}^d}  (\hatn ( \tau_\vec{-k}\boldsymbol{\phi}) \cdot \bsigma) ~ \ket{\vec{k}}\bra{\vec{k}} ,
	\notag \\ & =
	\xc\sum_{\vec{k} \in \mathbb{Z}^d}  (\hatn (\boldsymbol{\phi} + \theta \vec{k}) \cdot \bsigma) ~ \ket{\vec{k}}\bra{\vec{k}},
\end{align}
and the translation of the Hamiltonian $H = H_t + H_\mathrm{xc}(\boldsymbol{\phi})$ can  be expressed in the compact, covariant form
\begin{align}
\hat{T}_\vec{m}	H( \boldsymbol{\phi} )\hat{T}_\vec{m}^\dagger
 = H( \tau_\vec{m}\boldsymbol{\phi} ) ,
\end{align}
or alternatively
\begin{align}
	\hat{T}_\vec{m}^\dagger	H( \boldsymbol{\phi} )\hat{T}_\vec{m}
	 = H( \boldsymbol{\phi} +  \theta\vec{m} ) .
	\end{align}
Combining the results above, the Hamiltonian can finally be cast into the form
\begin{equation}
	H =  \sum_{\vec{n} \in \mathbb{Z}^d} \hat{T}_\vec{n}  \sum_{\vec{m} \in \mathbb{Z}} h_\vec{n}(  \boldsymbol{\phi} + \theta \vec{m}  ) \ket{ \vec{m}} \bra{ \vec{m}} ,
\end{equation}
with the definition
\begin{equation}
	h_\vec{n}(  \boldsymbol{\phi} )
	\equiv
	\left\lbrace
	\begin{array}{ll}
		\xc (\hatn ( \boldsymbol{\phi}) \cdot \bsigma) , & \vec{n} = 0  \\
		t~\id_2, & \exists l \in \lbrace 1,\cdots,d \rbrace\colon \vec{n} = \pm\vec{e}_l \\
		0, & \text{otherwise.}
		\end{array}
	\right.
\end{equation}

\section{Derivation of the torus commutation relation}
\label{app:commutation_relations}

The covariant form of the Hamiltonian demonstrates that it fits into a generic form which combines the action of the translation operator with matrix-valued functions on the $r$-torus $T^r$. 
A continuous function $f\colon T^r \to \mathbb{C}$ can be decomposed into a Fourier series as
\begin{align}
	f( \boldsymbol{\phi} ) & = \sum_{\vec{n} }  f_{\vec{n} }~ e^{2 \pi i \boldsymbol{\phi} \cdot \vec{n} }
	\notag \\
	& = \sum_{\vec{n} }  f_{\vec{n} }~ e^{2 \pi i \phi_1 n_1 } \cdots e^{2 \pi i \phi_r n_r }
	\notag \\
	& = \sum_{\vec{n} }  f_{\vec{n} } ~( e^{2 \pi i \phi_1  } )^{n_1} \cdots ( e^{2 \pi i \phi_r  } )^{n_r} 
	\notag \\
	& \equiv \sum_{\vec{n} }  f_{\vec{n} } ~u_1^{n_1} \cdots u_r^{n_r} .
\end{align}
In other words, the algebra of continuous functions on the torus is generated by $u_k = e^{2 \pi i \phi_k  }$.
One can condense this result into the presentation 
\begin{equation}
	C(T^r) = \Braket{ u_1, \ldots, u_r ~|~ [u_i, u_j] = 0} .
\end{equation}
The commutation relation between the unit lattice translation $\tau_l$ and the Fourier factor $u_k$ can be derived as
\begin{align}
	\tau_l u_k &=
	\exp\lbrace 
	2 \pi i  (\tau_l \phi_k )
	\rbrace \tau_l
	\\ \notag 
	&=
	\exp\lbrace 
	2 \pi  i ( ( \phi_k - ( \vec{e}_l \cdot \boldsymbol{\theta}_k \mod 1) ) \mod 1 )
	\rbrace \tau_l
	\\ \notag 
	&=
	\exp\lbrace 
	2 \pi  i  ( \phi_k - ( \vec{e}_l \cdot \boldsymbol{\theta}_k \mod 1))
	\rbrace \tau_l
	\\ \notag  &= 
	\exp\lbrace 
	-2 \pi  i ( \vec{e}_l \cdot \boldsymbol{\theta}_k \mod 1)
	\rbrace u_k \tau_l
	\\ \notag  &= 
	\exp\lbrace 
	-2 \pi  i  (\vec{e}_l \cdot \boldsymbol{\theta}_k )
	\rbrace u_k \tau_l
	\\ \notag  &= 
	\exp\lbrace 
	-2 \pi  i  \theta_{kl} 
	\rbrace u_k \tau_l 
\end{align}
which leads to the relation presented in the manuscript.

\section{A more precise description of the noncommutative torus}
\label{app:operator_algebra}

{\it The following section is adapted from \cite{Liu2022}}. By defining $\boldsymbol{\alpha}=(\tau_1, \ldots, \tau_d, u_1,\ldots u_r )$, the commutation relations can be summarized to $\alpha_l \alpha_k = e^{2 \pi i \Theta_{lk}} \alpha_k  \alpha_l$, where
\begin{equation}
    \Theta = \begin{pmatrix}
    0 & -\theta^T \\
    \theta & 0
    \end{pmatrix} .
\end{equation}
The manuscript summarizes the observable algebra of a multi-$\vec{q}$ texture as the universal $C^\ast$-algebra given by the presentation
\begin{align}
    \mathcal{A}_\Theta &= \Braket{ \alpha_1, \cdots, \alpha_{\deff}
    |  \alpha_l \alpha_k = e^{ 2 \pi i \Theta_{lk}} \alpha_k  \alpha_l },
\end{align}
with $\deff = r+ d$.
$ \Theta_{lk}$ is considered as an antisymmetric  $\deff\times \deff$  matrix with entries from $\RM/\ZM$.
A generic element of the algebra can be presented in the form
\begin{align}
a & = \sum_{\vec{q} \in \ZM^{\deff}} a_{\vec{q}}
 \alpha_{\vec{q}},  \alpha_{\vec{q}} =\alpha_1^{q_1} \ldots\alpha_\deff^{q_\deff},  a_{\vec{q}} \in \mathrm{Mat}_{2\times2} (\mathbb{C})
\notag \\
& = 	\sum_{\vec{q} \in \ZM^{d}} a( \boldsymbol{\phi}, \vec{q}) 
\, 
\alpha_1^{q_1} \ldots\alpha_d^{q_d},
\end{align}
where $ a(\boldsymbol{\phi},\vec{q}) $  is a continuous function $T^r \times \mathbb{Z}^d \to \mathrm{Mat}_{2\times 2}(\mathbb{C})$ with compact support.
The noncommutative torus accepts the trace
\begin{equation}
	\mathcal{T} \Big(
		\sum_{\vec{q} \in \ZM^{\deff}} a_{\vec{q}}
\, \alpha_{\vec{q}}
	\Big)
	= \tr~ a_{\vec
	0} .
\end{equation}
We define a representation of the noncommutative torus $\pi_{\boldsymbol{\phi}} \colon \mathcal{A}_\Theta \to \mathcal{B}(\ell^2(\mathbb{Z}^d \otimes \mathbb{C}^2))$ via the matrix elements
\begin{align}
	\braket{\vec{q}, \alpha
	|
	\pi_{\boldsymbol{\phi}} (a)
	| \vec{q}', \beta
	}
	=
	a_{\alpha \beta} ( \tau_{-\vec{q}} \boldsymbol{\phi}, \vec{q}'-\vec{q}) .
\end{align}
Constructed in this way, the representation fulfills the covariance condition
\begin{equation}
	\hat{T}_{\vec{m}}
	\pi_{\boldsymbol{\phi}} (a)
	\hat{T}_{\vec{m}}^\dagger
	= \pi_{\tau_{\vec{m}}\boldsymbol{\phi}} (a) ,
\end{equation}
which we previously confirmed to hold for the Hamiltonian.
Additionally, an involution is defined by
\begin{equation}
	a^\ast( \boldsymbol{\phi}, \vec{q})
	= a( \tau_{-\vec{q}} \boldsymbol{\phi}, -\vec{q})^\dagger .
\end{equation}
The $C^\ast$-algebra associated to $\mathcal{A}_{\Theta}$ is then given by the completion with respect to the norm 
\begin{equation}
	\| a \| = \sup\limits_{ \boldsymbol{\phi} \in T^r} \| \pi_{\boldsymbol{\phi}} a \| .
\end{equation}

\section{Some general elements of K-theory}
\label{app:k_theory}

{\it The following section is adapted from \cite{Liu2022}}. The general goal of the K-theory of operator algebras is to supply all independent topological invariants that can be associated to projections and unitary elements of an algebra.
In particular, the $K$-theory group $K_0(\mathcal{A}_\Theta)$ classifies the projections
\begin{equation}
p \in \Mm_\infty \otimes \Aa_\Theta, \quad p^2 = p^\ast=p,
\end{equation}
with respect to the von~Neumann equivalence relation
\begin{equation}\label{Eq-EquivRelation}
p \sim p' \quad \mbox{iff}  \quad p=vv' \ \  {\rm and} \ \ p' = v'v, 
\end{equation}
for some partial isometries $v$ and $v'$ with  $ vv', v'v \in \Mm_\infty \otimes \Aa_\Theta$. 
$\Mm_N$ is the algebra of $N \times N$ matrices with complex entries and $M_\infty$ is the direct limit of these algebras. 
For any $p$ from $\Mm_\infty \otimes \Aa_\Theta$, there exists $N \in \NM$ such that $p \in \Mm_N \otimes \Aa_\Theta$, hence we do not really need to work with infinite matrices. However, $\Mm_N$ can be canonically embedded into $\Mm_\infty$ and this convenient, because it enables $N$ to take flexible values.
There are two further equivalence relations for projections which could be used, and which lead to the same group $K_0(\mathcal{A}_\Theta)$ \cite[p.~18]{Park2008}:
\begin{itemize}
\item  Similarity equivalence:
\begin{equation}
p \sim_u p' \quad {\rm iff} \quad p'= u p u^\ast
\end{equation}
for some unitary element $u$ from $\Mm_\infty \otimes \Aa_\Theta$;
\item Homotopy equivalence:
\begin{equation}
p \sim_h p' \quad  {\rm iff} \quad \bm p(0)=p \ \  {\rm and} \ \  \bm p(1) = p'
\end{equation}
for some continuous function $\bm p : [0,1] \rightarrow \Mm_\infty \otimes \Aa_\Theta$, which always returns a projection. 
\end{itemize}
Homotopy equivalence is the topological equivalence as understood by condensed matter physicists. 
The equivalence class of a projection $p$ will be denoted by $[p]$, i.e., $[p]$ is the set
\begin{equation}
[p]= \big \{p' \in \Mm_\infty \otimes \Aa_\Theta \, ,  \ p' \sim p \big \}.
\end{equation}
If $p \in \Mm_N \otimes \Aa_\Theta$ and $p' \in \Mm_M \otimes \Aa_\Theta$ are two projections, then $\begin{pmatrix} p & 0 \\ 0 & p' \end{pmatrix}$ is a projection from $\Mm_{N+M} \otimes \Aa_\Theta$ and one can define the addition
\begin{equation}
[p] \oplus [p'] = \left [  \left( \begin{matrix} p & 0 \\ 0 & p' \end{matrix} \right) \right ],
\end{equation}
which provides a semigroup structure on the set of equivalence classes. Then $K_0(\Aa_\Theta)$ is its  enveloping group \cite{Blackadar1998} and, for the noncommutative $\deff$-torus,
\begin{equation}\label{Eq:K0} 
K_0(\Aa_\Theta) = \ZM^{2^{\deff-1}},
\end{equation}
regardless of $\Theta$ and where $\deff = r+d$.
As such, there are $2^{\deff-1}$ generators $[E_J]$, which can be uniquely labeled by the subsets of indices $J \subseteq \{1,\ldots,d\}$ of even cardinality \cite{Prodan2016}. Eq.~\eqref{Eq:K0} assures us that, for any projection $p$ from $\Mm_\infty \otimes \Aa_\Theta$, one has
\begin{equation}\label{Eq-GenExpansion}
[p]  = \sum_{J \subseteq \{1,\ldots,\deff \}}^{|J|={\rm even}} n_J \, [E_J], 
\end{equation} 
where the coefficients $n_J$ are integer numbers that do not change as long as $p$ is deformed inside its $K_0$-class. 
Specifically, two homotopically equivalent projections will display the same coefficients, hence $\{n_J\}_{|J|={\rm even}}$ represent the {\it complete} set of topological invariants associated to the projection $p$. 
Furthermore, two projections that display the same set of coefficients are necessarily in the same $K_0$-class. 

\section{Differential calculus on the noncommutative torus}
\label{app:differential_calculus}

As a preliminary step to the calculation of Chern numbers on the noncommutative torus, a differential calculus needs to be established.
Let $\lambda_i \in \CM$, $|\lambda_i| =1$ and observe that commutation relations of $\mathcal{A}_\Theta$ are invariant with respect to:
\begin{equation}
\alpha_j \mapsto \lambda_j \alpha_j.
\end{equation}
As such, we can define a $\deff$-torus action:
\begin{equation}
\TM^\deff \ni \bm \lambda=(\lambda_1,\ldots,\lambda_\deff) \mapsto \rho_{\bm \lambda}: \mathcal{A}_\Theta \to \mathcal{A}_\Theta
\end{equation}
where the latter is the algebra automorphism:
\begin{align}
	A =& \sum_{\vec{q} \in \ZM^{\deff}} a_{\vec{q}}
	\, \alpha_1^{q_1} \ldots\alpha_\deff^{q_\deff}
    \notag \\
 \mapsto & 
	\sum_{\vec{q} \in \ZM^{\deff}} a_{\vec{q}}
	\, 
	\lambda_1^{q_1} \ldots\lambda_\deff^{q_\deff}
	\alpha_1^{q_1} \ldots\alpha_\deff^{q_\deff} .
\end{align}
Then the generators of the torus action:
\begin{equation}
\partial_i(A) = i \partial_{\lambda_i} \rho_{\bm \lambda}(A)|_{\bm \lambda \rightarrow 1}
=\sum_{\vec{q} \in \ZM^{\deff}} i q_i a_{\vec{q}}
\, \alpha_1^{q_1} \ldots\alpha_\deff^{q_\deff}
\end{equation}
provide derivations on the noncommutative $\deff$-torus. 
We again define our indices with respect to the index set $\mathcal{I} = \lbrace \tau_1 \cdots, \tau_d, u_1  \cdots, u_r  \rbrace$.
Since
\begin{align}
	\partial_{\phi_k}  e^{2\pi i \boldsymbol{\phi} \cdot \vec{n}} = 2\pi  i n_k  e^{2 \pi i \boldsymbol{\phi} \cdot \vec{n}},
\end{align}
one finds that the $u$-derivations are just given by the partial derivatives
\begin{align}
	\partial_{u_k}  A 
	=  (2\pi)^{-1}\partial_{\phi_k}  A. 
\end{align}
For the $\tau$-derivations, the representation on the Hilbert space evaluates to
\begin{align}
	\pi_{\boldsymbol{\phi}} (\partial_{\tau_k} A) = i [ \hat{X}_k, \pi_{\boldsymbol{\phi}} (A)],
\end{align}
where $\hat{\vec{X}} = \sum_{\vec{q}\in \mathbb{Z}^d} \vec{x}_{\vec{q}}\ket{\vec{q}} \bra{\vec{q}}$ is the position operator on the Hilbert space.

\section{Relation to Berry curvature}
\label{app:berry_curvature}

If the multi-$\vec{q}$ texture is commensurate with the lattice, a Bloch basis can be chosen.
We introduce the new basis notation for the orbital wave functions
\begin{equation}
	\ket{\vec{R}, \vec{q}, \alpha} = \ket{\vec{R}+\vec{x}_{\vec{q}}, \alpha} ,
\end{equation}
here $\vec{R}$ describes the lattice of the superstructure.
The lattice Fourier transform (Wannier basis) is given by:
\begin{equation}
	\ket{\vec{R}, \vec{q}, \alpha} = \frac{1}{\sqrt{N}}
	\sum_{\vec{k} \in \mathrm{1. BZ}} e^{-i \vec{k}\cdot\vec{R}} 
	\ket{\psi_{\vec{k} \vec{q} \alpha}},
\end{equation}
where $N$ is now the number of primitive cells in the system.
Let $\hat{A}$ now represent a translationally invariant operator (w.r.t. the superstructure), i.e.,
\begin{align}
	\hat{A}  = &\sum_{\vec{R}, \vec{R}'} \sum_{\vec{q}, \vec{q}' , \alpha, \beta}
	A_{\vec{R}-\vec{R}'}^{ \alpha, \beta, \vec{q}, \vec{q}'} \ket{\vec{R}, \vec{q}, \alpha} \bra{\vec{R}',\vec{q}', \beta}
	\notag \\
	=& \frac{1}{N} \sum_{\vec{k}, \vec{k}' \in \mathrm{1. BZ}}
	\sum_{\vec{R}, \vec{R}'} \sum_{\vec{q}, \vec{q}' , \alpha, \beta}
	A_{\vec{R}-\vec{R}'}^{ \alpha, \beta, \vec{q}, \vec{q}'} 
    \notag \\ & \times
	e^{-i \vec{k}\cdot\vec{R}} 
	e^{+i \vec{k}'\cdot\vec{R}'} 
	\ket{\psi_{\vec{k} \vec{q} \alpha}} \bra{\psi_{\vec{k}' \vec{q}' \beta}}
	\notag \\
	= & \sum_{\vec{k}, \vec{k}' \in \mathrm{1. BZ}}
	 \sum_{\vec{q}, \vec{q}' , \alpha, \beta}
	A_{\vec{k}, \vec{k}'}^{ \alpha, \beta, \vec{q}, \vec{q}'} 
	\ket{\psi_{\vec{k} \vec{q} \alpha}} \bra{\psi_{\vec{k}' \vec{q}' \beta}},
\end{align}
where 
\begin{align}
	A_{\vec{k}, \vec{k}'}^{ \alpha, \beta, \vec{q}, \vec{q}'} 
	& =  \frac{1}{N}
	\sum_{\vec{R}, \vec{R}'}A_{\vec{R}-\vec{R}'}^{ \alpha, \beta, \vec{q}, \vec{q}'} 
	e^{-i \vec{k}\cdot\vec{R}} 
	e^{+i \vec{k}'\cdot\vec{R}'} 
	\notag \\
	& =  \frac{1}{N}
	\sum_{\vec{R}, \vec{R}'}A_{\vec{R}}^{ \alpha, \beta, \vec{q}, \vec{q}'} 
	e^{-i \vec{k}\cdot\vec{R}} 
	e^{-i \vec{k}\cdot\vec{R}'} 
	e^{+i \vec{k}'\cdot\vec{R}'} 
	\notag \\
	& =   \delta_{\vec{k}, \vec{k}'}
	\sum_{\vec{R}}A_{\vec{R}}^{ \alpha, \beta, \vec{q}, \vec{q}'} 
	e^{-i \vec{k}\cdot\vec{R}} 
	\notag \\
	& \equiv   \delta_{\vec{k}, \vec{k}'}
	(A_{\vec{k}})_{ \alpha, \beta, \vec{q}, \vec{q}'} .
\end{align}

This means that the trace of any operator product of translationally invariant operators is given by
\begin{align}
\mathcal{T} \left( \hat{A}^1 \cdots  \hat{A}^j \right) &=
\frac{1}{V}
\lim\limits_{N\to \infty}	\frac{1}{N} \sum_{\vec{k} \in \mathrm{1.BZ}} ~ \tr~ A^1_{\vec{k}} \cdots A^j_{\vec{k}}
\notag \\
& =   \int\limits_{\mathrm{1.BZ}}  \frac{\dd^d \vec{k}}{(2\pi)^d}~ \tr~ A^1_{\vec{k}} \cdots A^j_{\vec{k}},
\end{align}
where $V$ is the volume of the primitive unit cell and the trace $\tr$ includes the internal lattice degrees of freedom within the unit cell (in the addition to the spin degree).
Take now a covariant operator
\begin{align}
	\hat{A} &=
	\sum_{\vec{R}} \sum_{\alpha\beta}
	\sum_{\vec{q} }  A_{\alpha,\beta}( \tau_\vec{-q}\boldsymbol{\phi})~ \ket{\vec{R},\vec{q}, \alpha}\bra{\vec{R},\vec{q}, \beta} ,
\end{align}
and therefore
\begin{align}
	(A_{\vec{k}})_{ \alpha, \beta, \vec{q}, \vec{q}'}
	= \delta_{\vec{q}, \vec{q}'} (A_{\vec{k}}( \tau_\vec{-q}\boldsymbol{\phi}))_{ \alpha, \beta} . 
\end{align}
We split the trace in two parts $\tr = \tr_{\vec{q}} \tr_{\sigma} $ according to the atomistic degrees of freedom and the spin degree of freedom. 
By carrying out the operator product of covariant operators, one finds
\begin{align}
	&\mathcal{T} \left( \hat{A}^1 \cdots  \hat{A}^j \right)  =   \int\limits_{\mathrm{1.BZ}}\frac{\dd^d \vec{k}}{(2\pi)^d}~ \tr_{\vec{q}} \tr_\sigma ~ A^1_{\vec{k}} \cdots A^j_{\vec{k}}
	\notag \\
	&=
 \sum_{\vec{q}}\int\limits_{\mathrm{1.BZ}} \frac{\dd^d \vec{k}}{(2\pi)^d}~ \tr_\sigma ~ A^1_{\vec{k}}( \tau_\vec{-q}\boldsymbol{\phi})\cdots A^j_{\vec{k}}( \tau_\vec{-q}\boldsymbol{\phi})
	\notag \\
	&  
	\to
	\int\limits_{\mathrm{1.BZ}} \frac{\dd^d \vec{k}}{(2\pi)^d} \int \limits_{\Omega}\dd^r \boldsymbol{\phi}~ \tr_\sigma ~ A^1_{\vec{k}}( \boldsymbol{\phi})\cdots A^j_{\vec{k}}( \boldsymbol{\phi}) .
	\end{align}
Here, the limit $\to$ indicates the transition to a smooth magnetic texture, which is supported by a larger and larger amount of atomic sites in the primitive cell of the superstructure.
As a further ingredient, one needs that the action of the translation operator is ergodic on $\Omega$ in the smooth limit.

Assuming $\hat{A}$ is diagonal in $\vec{q}$ (as is the case for the covariant operators):
\begin{align}
	&i [ \hat{X}_i, A]
	 =  \sum_{\vec{R}, \vec{R}'} \sum_{\vec{q} , \alpha, \beta}
	i(\vec{R}-\vec{R}')_i A_{\vec{R}-\vec{R}'}^{ \alpha, \beta, \vec{q}} \ket{\vec{R}, \vec{q}, \alpha} \bra{\vec{R}',\vec{q}, \beta}
	\notag \\
	& =    \sum_{\vec{k} \in \mathrm{1. BZ}} \sum_{\vec{q} , \alpha, \beta}
	\sum_{\vec{R}}   i R_i~ A_{\vec{R}}^{ \alpha, \beta, \vec{q}} 
	e^{-i \vec{k}\cdot\vec{R}} 
	\ket{\psi_{\vec{k} \vec{q} \alpha}} \bra{\psi_{\vec{k} \vec{q} \beta}}
	\notag \\
	& =   -\sum_{\vec{k} \in \mathrm{1. BZ}} \sum_{\vec{q} , \alpha, \beta}\partial_{k_i}
	\sum_{\vec{R}}   A_{\vec{R}}^{ \alpha, \beta, \vec{q}} 
	e^{-i \vec{k}\cdot\vec{R}} 
	\ket{\psi_{\vec{k} \vec{q} \alpha}} \bra{\psi_{\vec{k} \vec{q} \beta}}
	\notag \\
	& =   \sum_{\vec{k} \in \mathrm{1. BZ}}\sum_{\vec{q} , \alpha, \beta}
	(-\partial_{k_i}A_{\vec{k}})_{ \alpha, \beta, \vec{q}}
	\ket{\psi_{\vec{k} \vec{q} \alpha}} \bra{\psi_{\vec{k} \vec{q} \beta}} 
\end{align}
For covariant operators, we therefore have the correspondence dictionary for the covariant Bloch representation
\begin{align}
	\pi_{\boldsymbol{\phi}}(A)
	& \to A_{\vec{k}}(\boldsymbol{\phi})
	\label{eq:rep1}
\\
	\pi_{\boldsymbol{\phi}}(\partial_{u_j} A)
	& 
	\to \partial_{\phi_j} A_{\vec{k}}(\boldsymbol{\phi}) / (2\pi)
	\label{eq:rep2}
	\\
	\pi_{\boldsymbol{\phi}}(\partial_{\tau_j} A)
	& 
	\to -\partial_{k_j} A_{\vec{k}}(\boldsymbol{\phi}) 
	\label{eq:rep3}
\\
 \mathcal{T} & \to  \int\limits_{\mathrm{1.BZ}} \frac{\dd^d \vec{k}}{(2\pi)^d}~ \tr_\sigma ~ \sum_{\vec{q}}\tau_{\vec{q}} \rhd,
\end{align}
where $\tau_{\vec{q}} \rhd$ denotes the action: 
\begin{equation}
	\tau_{\vec{q}} \rhd A^1( \boldsymbol{\phi}) \cdots  A^j( \boldsymbol{\phi}) \equiv A^1( \tau_{-\vec{q}}\boldsymbol{\phi}) \cdots  A^j( \tau_{-\vec{q}}\boldsymbol{\phi})
\end{equation}
And, in the limit of smooth textures and ergodic action,
\begin{align}
	\sum_{\vec{q}}\tau_{\vec{q}} \rhd \to \int\limits_{\Omega }\dd^r \boldsymbol{\phi} .
\end{align}

Now that the differential calculus on the torus is established, the Chern numbers can be defined.
The Chern number of a projection $P$ to gap $g$ and associated to a subset of indices $J$ of even cardinality is given by
\begin{equation}
	    {\rm Ch}_{J'}(g) = \frac{(2\pi i)^{|J'|/2}}{( |J'|/2 )!} \sum_{\sigma \in \Ss_{|J'|}} (-1)^\sigma \Tt\Big ( P \prod_{j \in J'} \partial_{\sigma_j}P \Big ) ,
\end{equation}
where for $J = \emptyset$, we define ${\rm Ch}_\emptyset(P) = \Tt(P)$.
The structure of the noncommutative torus imposes relations on the Chern numbers.
These can be found by studying the values of the Chern numbers on the $K_0$-generators of $\mathcal{A}_\Theta$, which can be found in \cite{Prodan2016}[p.~141]:
\begin{equation}\label{Eq:ChernValues}
{\rm Ch}_{J'} [E_J] = \left \{ 
\begin{array}{l}
0 \ {\rm if} \ J'\nsubseteq J  , \\
1 \ {\rm if} \ J' = J , \\
{\rm Pf}(\Theta_{J\setminus J'}) \ {\rm if} \ J' \subset J,
\end{array}
\right .  
\end{equation}
where $ J, J' \subset \{1,\ldots,\deff\}$.
Since the Chern numbers are also linear maps, their values on the gap projection $[P_G] = \sum_{J} n_J \, [e_J]$ can be straightforwardly computed from Eq.~\eqref{Eq:ChernValues}:
\begin{equation}\label{Eq:ChernVal}
{\rm Ch}_{J'}(g) = n_{J'}(g) + \sum_{J' \subsetneq J} n_J(g) \, {\rm Pf}(\Theta_{J\setminus J'}).
\end{equation} 
The $K$-theory of the noncommutative torus therefore imposes relations among the various Chern numbers. 
Top Chern number corresponding to $J'=\{1,\ldots,\deff\}$ is always an integer, but the lower Chern numbers may not be.

To illustrate the special case of a commensurate texture, consider the special case of $d=r=2$ and $\deff =d+r=4$.
Via the correspondence dictionary, we find the top Chern number provided by the expression (for $J=\lbrace \tau_1, \tau_2, u_1, u_2 \rbrace $)
\begin{align}
{\rm Ch}_{J}(g)
= & -\frac{1}{2}
\int\limits_{\mathrm{1.BZ}} \frac{\dd^d \vec{k}}{(2\pi)^d}~\sum_{\vec{q}}\tau_{\vec{q}} \rhd\sum_{\sigma \in \Ss_{4}} (-1)^\sigma 
\notag \\ & 
\times \tr_\sigma ~ 
P_{\vec{k}}(\boldsymbol{\phi}) \prod_{j \in J} \partial_{\sigma_j}P_{\vec{k}}(\boldsymbol{\phi}),
\end{align}
where the representations of Eq.~\ref{eq:rep1} - Eq.~\ref{eq:rep2} have already been inserted.
We identify the Berry curvature 
\begin{equation}
	F_{\sigma_1, \sigma_2}(\vec{k}, \boldsymbol{\phi})
	= i P_{\vec{k}}(\boldsymbol{\phi}) [
	\partial_{\sigma_1} P_{\vec{k}}(\boldsymbol{\phi}), \partial_{\sigma_2} 
	P_{\vec{k}}(\boldsymbol{\phi}) ]
\end{equation}
and write
\begin{widetext}
\begin{align}
\sum_{\sigma \in \Ss_{4}} (-1)^\sigma \tr_\sigma ~ 
P_{\vec{k}}(\boldsymbol{\phi}) \prod_{j \in J} \partial_{\sigma_j}P_{\vec{k}}(\boldsymbol{\phi}) &=
\epsilon^{\alpha\beta\gamma\delta}
\tr_\sigma ~
P_{\vec{k}}(\boldsymbol{\phi}) \partial_{\sigma_\alpha }P_{\vec{k}}(\boldsymbol{\phi})
\partial_{\sigma_\beta }P_{\vec{k}}(\boldsymbol{\phi})
\partial_{\sigma_\gamma }P_{\vec{k}}(\boldsymbol{\phi})
\partial_{\sigma_\delta }P_{\vec{k}}(\boldsymbol{\phi})
\notag \\
&=
\epsilon^{\alpha\beta\gamma\delta}
\tr_\sigma ~
P_{\vec{k}}(\boldsymbol{\phi}) \partial_{\sigma_\alpha }P_{\vec{k}}(\boldsymbol{\phi})
\partial_{\sigma_\beta }P_{\vec{k}}(\boldsymbol{\phi})
P_{\vec{k}}(\boldsymbol{\phi}) 
\partial_{\sigma_\gamma }P_{\vec{k}}(\boldsymbol{\phi})
\partial_{\sigma_\delta }P_{\vec{k}}(\boldsymbol{\phi})
\notag \\
& = - \frac{1}{4}
\epsilon^{\alpha\beta\gamma\delta}
\tr_\sigma ~F_{\alpha \beta}(\vec{k}, \boldsymbol{\phi}) F_{\gamma\delta}(\vec{k}, \boldsymbol{\phi}) .
\end{align}
\end{widetext}   
\vfill
Inserting this result into the expression for the Chern number gives 
\begin{align}
	&{\rm Ch}_{J}(g) = \frac{1}{8 }\sum_{\vec{q}}
    \int\limits_{\mathrm{1.BZ}} \frac{\dd^d \vec{k}}{(2\pi)^d}~ \epsilon^{\alpha\beta\gamma\delta}
    \notag \\
    & \phantom{{\rm Ch}_{J}(g) = } \times
 \tr_\sigma F_{\alpha \beta}(\vec{k}, \tau_{-\vec{q}}\boldsymbol{\phi}) F_{\gamma\delta}(\vec{k}, \tau_{-\vec{q}}\boldsymbol{\phi}) 
 \notag \\
 & = \frac{1}{32 \pi^2}\sum_{\vec{q}}
 \int\limits_{\mathrm{1.BZ}}\dd^d \vec{k} ~ \epsilon^{\alpha\beta\gamma\delta}
 \tr_\sigma F_{\alpha \beta}(\vec{k}, \tau_{-\vec{q}}\boldsymbol{\phi}) F_{\gamma\delta}(\vec{k}, \tau_{-\vec{q}}\boldsymbol{\phi}) 
\end{align}
Taking the limit of smooth textures of this expression, one then obtains
\begin{align}
 & \to \frac{1}{32 \pi^2} \int\limits_{\Omega} \dd^r\boldsymbol{\phi}
 \int\limits_{\mathrm{1.BZ}} \dd^d \vec{k} ~ \epsilon^{\alpha\beta\gamma\delta}
 \tr_\sigma F_{\alpha \beta}(\vec{k},\boldsymbol{\phi}) F_{\gamma\delta}(\vec{k}, \boldsymbol{\phi}) 
 \notag \\
 & = \frac{1}{32 \pi^2} \int\limits_{T^{\deff}}  \dd^\deff \boldsymbol{\lambda}
  ~ \epsilon^{\alpha\beta\gamma\delta}
 \tr_\sigma F_{\alpha \beta}(\boldsymbol{\lambda}) F_{\gamma\delta}(\boldsymbol{\lambda}) ,
\end{align}
which is the familiar expression for the second Chern number in terms of the Berry curvature~\cite{Qi2008}.
Repeating the same calculation for the case of $d=r=1$ and $\deff =d+r=2$, with $J=\lbrace \tau u \rbrace $, one finds
\begin{align}
	{\rm Ch}_{J}(g)
 & =- \frac{1}{2 \pi}\sum_{\vec{q}}
 \int\limits_{\mathrm{1.BZ}}\dd^d \vec{k} ~
 \tr_\sigma ~F_{\tau u}(\vec{k}, \tau_{-\vec{q}}\boldsymbol{\phi}) 
 \notag \\
 & \to -\frac{1}{2 \pi} \int\limits_{\Omega} \dd^r\boldsymbol{\phi}
 \int\limits_{\mathrm{1.BZ}} \dd^d \vec{k} ~
 \tr_\sigma ~F_{\tau u}(\vec{k},\boldsymbol{\phi}) ) 
 \notag \\
 & = -\frac{1}{2 \pi} \int\limits_{T^{\deff}}  \dd^\deff \boldsymbol{\lambda}
  ~
 \tr_\sigma ~F_{\tau u}(\boldsymbol{\lambda})  ,
\end{align}
which, in this case, is representing the usual expression for the first Chern number in terms of the Berry curvature~\cite{Qi2008}.

\section{The \texorpdfstring{$\Theta$-matrix}{Theta-matrix} for \texorpdfstring{3$\vec{q}$}{3q} states on the triangular lattice}
\label{app:theta_matrix_triangular_lattice}

In this section, we discuss the construction of the skyrmion $3\vec{q}$-state on the triangular lattice as it appears in the manuscript.
Real- and reciprocal space lattice vectors are introduced via
\begin{align}
	\vec{a}_1 &= (1,0)^T
	\\
	\vec{a}_2 &= (1/2,\sqrt{3}/2 )^T
	\\
	\vec{b}_1 &= 2 \pi (1, -1/\sqrt{3})^T
	\\
	\vec{b}_2 &= 2 \pi (0, 2 / \sqrt{3})^T .
\end{align}
With respect to these lattice vectors, the $\vec{q}$-vectors of the texture are given by
\begin{align}
	\vec{q_1} &= \theta_1 \vec{b}_2 \\
	\vec{q_2} &= \theta_1\vec{b}_1 \\ 
	\vec{q_3} &= \theta_1(-\vec{b}_1- \vec{b_2}) 
\end{align}
One can confirm that these vectors form the vertices of an equilateral triangle and that $\sum_i \vec{q}_i = 0$.
From the definition follows that the $\theta$-matrix is given by
\begin{equation}
	\theta =
	\theta_1 
	\begin{pmatrix}
		0 & 1 \\
		1 & 0 \\
		-1 & -1
	\end{pmatrix} .
\end{equation}
As initial phases we take $\boldsymbol{\phi} = (0,0,\pi)$.  
The respective Chern number decomposition can be found in table \ref{tab:3q_2d} at the end of this document (the analogous case for a 2$\vec{q}$-state in $d=1$ and $d=2$ is shown in \ref{tab:2q_1d} and  \ref{tab:2q_2d} respectively).
Let $R^z_{2\pi/3}$ represent a $-2\pi/3$ rotation around the $z$-axis.
Then we write
\begin{widetext}
\begin{align}
	\hatn_\mathrm{SkX}(\vec{x}) &=  \sum_{i=1}^{3 } (R^z_{2\pi/3})^{i-1} \hatn_\mathrm{hx} ( ((R^z_{2\pi/3})^{i-1} \vec{q}_1 )\cdot \vec{x}/ (2\pi) + \phi_i ) 
	\\
	\hatn_\mathrm{XY-V}(\vec{x}) &= \sum_{i=1}^{3 } (R^z_{2\pi/3})^{i-1} \hatn_\mathrm{sdw} ( ((R^z_{2\pi/3})^{i-1} \vec{q}_1 )\cdot \vec{x}/ (2\pi)  + \phi_i ) .
\end{align}
\end{widetext}
Here, the skyrmion lattice $\hatn_\mathrm{SkX}$ is therefore constructed from a coherent superposition of three spin helices (hx), and the vortex lattice $\hatn_\mathrm{XY-V}$ is constructed  from a coherent superposition of spin density waves (sdw).
Respectively, these are defined by
\begin{align}
\hatn_\mathrm{hx} (\psi) &=  (0, \sin(\psi), \cos(\psi))^T,
\\
\hatn_\mathrm{sdw} (\psi) &=  (\sin(\psi),0,0)^T .
\end{align}
For the SkX is state, the result of the formula is always normalized by
$\hatn_\mathrm{SkX}(\vec{x}) \to \hatn_\mathrm{SkX}(\vec{x}) / \| \hatn_\mathrm{SkX}(\vec{x})\| $, while for the $XY-V$ state, one scales the result such that
\begin{equation}
	\sup_\vec{x} \| 	\hatn_\mathrm{XY-V}(\vec{x}) \| = 1.
\end{equation}	
As the exact diagonalization of the Hamiltonian is computationally more demanding in $d=2$ dimensions compared to the $d=1$ case, we combine the spectra of different linear system sizes $N\in[19, 79]$ (i.e. there are $N$ lattice unit cells per dimension).
The $\theta_1$ are sampled again at rational values $\theta_1 = m /N $ with $m \in \mathbb{Z}$ and $ 0\leq m \leq N$.
Since $N$ is not necessarily prime, some $\theta_1$-values would be sampled multiple times.
When this occurs for two different values of $N$, we always choose the larger system size to obtain a better spectral resolution.

\section{Discussion of the relation to emergent magnetic fields}
\label{app:emergent_magnetic_fields}

In the adiabatic limit of smooth textures and strong exchange coupling, our theory should reduce to the well-known language of emergent magnetic fields.
To discuss the adiabatic limit, we introduce the unitary transformation
\begin{equation}
	U^\dagger(\vec{x}) ( \hatn(\vec{x})\cdot \bsigma ) U(\vec{x})  = \sigma_z .
\end{equation}
By parameterizing the magnetization vector in polar coordinates $\hatn = \hatn(\theta, \phi)$ in spherical coordinates, this transformation can be formulated explicitly as $\mathcal{U} = \hatn(\theta/2, \phi)\cdot \bsigma \equiv \vec{m} \cdot \bsigma$.
The discretization on the lattice is given by
\begin{equation}
	U( \hat{\vec{x}}) = \sum_{\vec{k} \in \mathbb{Z}^d} U( \vec{x}_{\vec{k}} )\ket{\vec{k}} \bra{\vec{k}} .
\end{equation}
Applying the transformation to the Hamiltonian, one finds
\begin{align}
	U( \hat{\vec{x}})^\dagger H  U( \hat{\vec{x}}) = \sum_{\braket{\vec{k},\vec{l}} \in \mathbb{Z}^{2d}} t_{\vec{k}\vec{l}} \ket{\vec{k}}\bra{\vec{l}}
	+ \xc\sum_{\vec{k} \in \mathbb{Z}^d}  \sigma_z ~ \ket{\vec{k}}\bra{\vec{k}} ,
\end{align}
where $t_{\vec{k}\vec{l}} = t U^\dagger(\vec{x}_\vec{k}) U(\vec{x}_\vec{l})$.
In the limit $\xc / t \to \infty$, one can project onto the eigenstates $\sigma = \pm 1$ of $\sigma_z$ in order to arrive at the effective Hamiltonian
\begin{equation}
	H_\mathrm{eff}^\sigma
	= \sum_{\braket{\vec{k},\vec{l}} \in \mathbb{Z}^{2d}} t_{\vec{k}\vec{l}, \sigma}^{\mathrm{eff}} \ket{\vec{k}}\bra{\vec{l}},
\end{equation}
where
$
t_{\vec{k}\vec{l}, \sigma}^{\mathrm{eff}} =t  \braket{\sigma |  U^\dagger(\vec{x}_\vec{k}) U(\vec{x}_\vec{l})  | \sigma }
$.
In the continuous case, a vector potential can be defined as
$
A_i = - i \hbar U^\dagger \partial_i U  / e ,
$
which has the adiabatic projection
\begin{equation}
	A_i^{\pm} = \pm \frac{\hbar}{e} ( \vec{m} \times \partial_i \vec{m} )_z .
\end{equation}
From this, one obtains the emergent magnetic field
\begin{equation}
	B_z^\pm 
	= (\nabla \times \vec{A}^\pm)_z = \pm \frac{\hbar}{2 e} \hatn \cdot (\partial_x \hatn \times \partial_y \hatn) .
\end{equation}
For an isolated skyrmion of topological charge 
\begin{equation}
	\mathcal{Q} = \frac{1}{4 \pi}\int\limits_{\mathbb{R}^2} \dd \vec{x}~ \hatn \cdot (\partial_x \hatn \times \partial_y \hatn) \in \mathbb{Z}
\end{equation}
is quantized.
The emergent flux in this case is
\begin{align}
	\Phi^\pm & = \int\limits_{\mathbb{R}^2} \dd \vec{x}~ B_z^\pm 
    \notag \\
	& =\pm \frac{\hbar}{2 e} \int\limits_{\mathbb{R}^2} \dd \vec{x}~ \hatn \cdot (\partial_x \hatn \times \partial_y \hatn)
    \notag \\
	&=\pm 2 \pi \frac{\hbar}{ e}  \mathcal{Q} .
\end{align}

Applying the translation operator to the previously defined unitary operator, we find
\begin{align}
	\hat{T}_\vec{m} U( \hat{\vec{x}}) \hat{T}_\vec{m}^\dagger
	& = \sum_{\vec{k}} U(\vec{x}_\vec{k}) \ket{\vec{k}+\vec{m}} \bra{\vec{k}+\vec{m}}
	\notag \\
	& = \sum_{\vec{k}} U( \vec{x}_{\vec{k}- \vec{m}} ) \ket{\vec{k}} \bra{\vec{k}}
	\notag \\
	& = U( \hat{\vec{x}} - \vec{x}_\vec{m}),
\end{align}
from which one can obtain the relation
$\hat{T}_\vec{m} U( \hat{\vec{x}}) = U( \hat{\vec{x}} - \vec{x}_\vec{m}) \hat{T}_\vec{m}$.
Within the changed frame of reference, the new unit translation operator is given by
\begin{align}
	\hat{S}_i &\equiv U^\dagger ( \hat{\vec{x}}) \hat{T}_i U ( \hat{\vec{x}})
	\notag \\ &=
	U^\dagger ( \hat{\vec{x}}) U( \hat{\vec{x}} - \vec{a}_i) \hat{T}_i  .
\end{align}

\begin{table}[t]
	\begin{tabular}{L|L}
		J' & \mathrm{Ch}_{J'} \\ \toprule
 \{\} & \theta _1 n_{\left\{\tau _1,u_1\right\}}+\theta _2 n_{\left\{\tau _1,u_2\right\}}+n_{\{\}} \\
\left\{\tau _1,u_1\right\} & n_{\left\{\tau _1,u_1\right\}} \\
\left\{\tau _1,u_2\right\} & n_{\left\{\tau _1,u_2\right\}} \\
\left\{u_1,u_2\right\} & n_{\left\{u_1,u_2\right\}} 
	\end{tabular}
\caption{Chern number expansion for a 2$\vec{q}$-state in $d=1$ dimensions with $\theta = (\theta_1, \theta_2)^T$ (e.g. the 2-$\vec{q}$ helicoids)
}
\label{tab:2q_1d}
\end{table}

We now assume that $\hatn$ is given by a multi-$\vec{q}$ state in $d=2$ dimensions, characterized by a single pitch variable $\theta_1$.
For a smoothly varying texture (limit of small $\theta_1$), the pre-factor can be expanded:
\begin{align}
	U^\dagger ( \vec{x}) U( \vec{x} - \vec{a}_i) 
	& = \id_2
	- U^\dagger ( \vec{x})( \vec{a}_i \cdot \nabla) U( \vec{x} ) + \mathcal{O}(\theta_1^2)
	\notag \\
	& = \id_2
	+ i  e~\vec{a}_i  \cdot \vec{A} / \hbar + \mathcal{O}(\theta_1^2)
	\notag \\
	& = \id_2
	+ \frac{i e}{\hbar} \int\limits_{\vec{x}}^{\vec{x}+\vec{a}_i} \dd\vec{r}~\vec{A}  + \mathcal{O}(\theta_1^2) 
	\notag \\
	& = \exp\left( \frac{i e}{\hbar} \int\limits_{\vec{x}}^{\vec{x}+\vec{a}_i} \dd\vec{r}~\vec{A} \right) + \mathcal{O}(\theta_1^2) ,
\end{align}
were we have implicitly used the adiabatic projection into a spin-subspace.
Introduce the shorthand notation
\begin{align}
	\uparrow_{\vec{x}}^{\vec{x}+\vec{a} } &\equiv \exp\left( \frac{i e}{\hbar} \int\limits_{\vec{x}}^{\vec{x}+\vec{a}_i} \dd\vec{r}~\vec{A} \right) ,
	\\
	\downarrow_{\vec{x}}^{\vec{x}+\vec{a} } &\equiv \exp\left( -\frac{i e}{\hbar} \int\limits_{\vec{x}}^{\vec{x}+\vec{a}_i} \dd\vec{r}~\vec{A} \right) .
\end{align}
Using this notation, one can derive the commutation relations
\begin{align}
	S_1 S_2 &=
	\uparrow_{\vec{x}}^{\vec{x}+\vec{a}_1 } T_1 \uparrow_{\vec{x}}^{\vec{x}+\vec{a}_2 } T_2 
	\notag \\
	&=
	\uparrow_{\vec{x}}^{\vec{x}+\vec{a}_1 } \uparrow_{\vec{x}-\vec{a}_1}^{\vec{x}+\vec{a}_2 -\vec{a}_1} T_2  T_1
	\notag \\
	&=
	\uparrow_{\vec{x}}^{\vec{x}+\vec{a}_1 } \uparrow_{\vec{x}-\vec{a}_1}^{\vec{x}+\vec{a}_2 -\vec{a}_1} T_2 \downarrow_{\vec{x}}^{\vec{x}+\vec{a}_1 } S_1
	\notag \\
	&=
	\uparrow_{\vec{x}}^{\vec{x}+\vec{a}_1 } \uparrow_{\vec{x}-\vec{a}_1}^{\vec{x}+\vec{a}_2 -\vec{a}_1} \downarrow_{\vec{x}+\vec{a}_2}^{\vec{x}+\vec{a}_1+\vec{a}_2 }T_2  S_1
	\notag \\
	&=
	\uparrow_{\vec{x}}^{\vec{x}+\vec{a}_1 } \uparrow_{\vec{x}-\vec{a}_1}^{\vec{x}+\vec{a}_2 -\vec{a}_1} \downarrow_{\vec{x}+\vec{a}_2}^{\vec{x}+\vec{a}_1+\vec{a}_2 }
	\downarrow_{\vec{x}}^{\vec{x}+\vec{a}_2 }
	S_2  S_1 .
\end{align}
Since
\begin{equation}
	\uparrow_{\vec{x}-\vec{a}_1}^{\vec{x}+\vec{a}_2 -\vec{a}_1}
	=
	\uparrow_{\vec{x}+\vec{a}_1}^{\vec{x}+\vec{a}_2 +\vec{a}_1} + \mathcal{O}(\theta_1^2),
\end{equation}
the combination of integrals amounts to clockwise line integral around the unit cell  anchored at $\vec{x}$.
We change this to a counter-clockwise orientation and apply the Stokes theorem to write the emergent flux as
\begin{equation}
	\Phi(\vec{x}) = \oint\limits_{\partial \mathrm{uc}(\vec{x})} \dd\vec{r}\cdot \vec{A} = \int\limits_{\mathrm{uc}(\vec{x})} \dd^2\vec{r}~ (\nabla \times \vec{A})_z .
\end{equation}              
We therefore find the commutation relation
\begin{equation}
	S_1 S_2 = e^{-i\hbar \Phi(\vec{x})/ e} S_2 S_1 + \mathcal{O}(\theta_1^2).
\end{equation}    
For the lattice of skyrmions with charge $\mathcal{Q}=1$, the emergent flux per magnetic unit cell is quantized, i.e., it is given by
$ | e\Phi_\mathrm{sk}/\hbar | = 2 \pi $.
On average, the flux per unit cell of the lattice is therefore given by
\begin{equation}
	\Braket{\Phi(\vec{x})}  =	\frac{2 \pi}{ \Braket{N_\mathrm{uc}}},
\end{equation}
where $\Braket{N_\mathrm{uc}}$ is the average number of lattice unit cells within a magnetic unit cell.
In $d=2$ dimensions, one has $\Braket{N_\mathrm{uc}} = 1 / \theta_1^2$
The algebra can then be approximated by replacing the exact flux $\Phi(\vec{x})$ per lattice unit cell by this average and one obtains the commutation relation
\begin{equation}
	S_1 S_2 \approx e^{-i 2\pi \theta_1^2} S_2 S_1 ,
\end{equation}
while at same time, $S_i$ commutes with the Fourier factors since the non-collinear magnetism has been transformed away.
All possible Chern numbers are then summarized by the table
\begin{align}
	\begin{tabular}{L|L}
		J' & \mathrm{Ch}_{J'} \\ \toprule
		\{\} & \theta _1^2 n_{\left\{s_1,s_2\right\}}+n_{\{\}} \\
		\left\{s_1,s_2\right\} & n_{\left\{s_1,s_2\right\}} 
	\end{tabular}
\end{align}
Consequently, the IDS in the gap $g$ for the effective system is given by the expansion
\begin{equation}
	\mathrm{IDS}(g) = n_\emptyset(g) + n_{s_1, s_2}(g) \theta_1^2 .
\end{equation}
By matching the coefficients, of the two limits one therefore finds
\begin{equation}
	n_{\lbrace s_1, s_2 \rbrace}(g) \sim n_{t^2 u^2}(g) , ~\text{for}~ | \xc / t| \to \infty, \theta_1 \to 0 .
\end{equation}
Further, the left-hand side can also be calculated directly as Chern number, since
\begin{equation}
	\mathrm{Ch}_{\lbrace s_1, s_2 \rbrace}(g) = n_{\lbrace s_1, s_2 \rbrace}(g) .
\end{equation}
Since the Chern number is invariant under unitary transformations of the Hamiltonian, this then leads to
\begin{equation}
	\mathrm{Ch}_{\lbrace t_1, t_2 \rbrace}(g) \sim \mathrm{Ch}_{\lbrace s_1, s_2 \rbrace}(g)\sim n_{t^2 u^2}(g) , 
\end{equation}
for $| \xc / t| \to \infty$ and $\theta_1 \to 0 $.
This means that the presence of a quantum anomalous Hall effect can be deduced from the IDS (where $n_{t^2 u^2}(g)$ can be extracted).
To rephrase this result: the relation holds, because we have shown that the physics of the asymptotic limit is described by a $2$-dimensional subalgebra of the full $(2+r)$-dimensional noncommutative torus generated by $	\hat{S}_i = \braket{\sigma |
U^\dagger ( \hat{\vec{x}}) U( \hat{\vec{x}} - \vec{a}_i) | \sigma } \hat{T}_i $.
This subalgebra is completely characterized by two topological integers $n_{\lbrace \rbrace}$ and $n_{\lbrace s_1, s_2 \rbrace}$ which can be directly extracted from the IDS.

\begin{table}[t]
\begin{tabular}{L|L}
J' & \mathrm{Ch}_{J'} \\ \toprule
\{\} & \theta _1^2 n_{\left\{\tau _1,\tau _2,u_1,u_2\right\}}+\theta _1 n_{\left\{\tau _1,u_2\right\}}+\theta _1 n_{\left\{\tau _2,u_1\right\}} \\
 & +n_{\{\}} \\
\left\{\tau _1,\tau _2\right\} & n_{\left\{\tau _1,\tau _2\right\}} \\
\left\{\tau _1,u_1\right\} & n_{\left\{\tau _1,u_1\right\}} \\
\left\{\tau _1,u_2\right\} & \theta _1 n_{\left\{\tau _2,u_1\right\}}+\theta _1 n_{\left\{\tau _1,\tau _2,u_1,u_2\right\}}+n_{\left\{\tau _1,u_2\right\}} \\
\left\{\tau _2,u_1\right\} & \theta _1 n_{\left\{\tau _1,u_2\right\}}+\theta _1 n_{\left\{\tau _1,\tau _2,u_1,u_2\right\}}+n_{\left\{\tau _2,u_1\right\}} \\
\left\{\tau _2,u_2\right\} & n_{\left\{\tau _2,u_2\right\}} \\
\left\{u_1,u_2\right\} & n_{\left\{u_1,u_2\right\}} \\
\left\{\tau _1,\tau _2,u_1,u_2\right\} & n_{\left\{\tau _1,\tau _2,u_1,u_2\right\}}
\end{tabular}
\caption{Chern number expansion for a 2$\vec{q}$-state in $d=2$ dimensions with $\theta = \theta_1 ((0,1),(1,0))$ (an example would be the 2-$\vec{q}$ skyrmion lattice).
}
\label{tab:2q_2d}
\end{table}
    
\clearpage

\hfuzz=280pt

\begin{turnpage}
	\begin{table}
	\begin{tabular}{L|L}
		J' & \mathrm{Ch}_{J'} \\ \toprule
		\{\} & \theta _1^2 n_{\left\{\tau _1,\tau _2,u_1,u_2\right\}}+\theta _1^2 n_{\left\{\tau
			_1,\tau _2,u_1,u_3\right\}}+\theta _1^2 n_{\left\{\tau _1,\tau
			_2,u_2,u_3\right\}}+\theta _1 n_{\left\{\tau _1,u_2\right\}}+\theta _1 n_{\left\{\tau
			_1,u_3\right\}}+\theta _1 n_{\left\{\tau _2,u_1\right\}}+\theta _1 n_{\left\{\tau
			_2,u_3\right\}}+n_{\{\}} \\
		\left\{\tau _1,\tau _2\right\} & n_{\left\{\tau _1,\tau _2\right\}} \\
		\left\{\tau _1,u_1\right\} & \theta _1 n_{\left\{\tau _2,u_3\right\}}+\theta _1
		n_{\left\{\tau _1,\tau _2,u_1,u_3\right\}}+n_{\left\{\tau _1,u_1\right\}} \\
		\left\{\tau _1,u_2\right\} & \theta _1 n_{\left\{\tau _2,u_1\right\}}+\theta _1
		n_{\left\{\tau _2,u_3\right\}}+\theta _1 n_{\left\{\tau _1,\tau
			_2,u_1,u_2\right\}}+\theta _1 n_{\left\{\tau _1,\tau
			_2,u_2,u_3\right\}}+n_{\left\{\tau _1,u_2\right\}} \\
		\left\{\tau _1,u_3\right\} & \theta _1 n_{\left\{\tau _2,u_1\right\}}+\theta _1
		n_{\left\{\tau _1,\tau _2,u_1,u_3\right\}}+n_{\left\{\tau _1,u_3\right\}} \\
		\left\{\tau _2,u_1\right\} & \theta _1 n_{\left\{\tau _1,u_2\right\}}+\theta _1
		n_{\left\{\tau _1,u_3\right\}}+\theta _1 n_{\left\{\tau _1,\tau
			_2,u_1,u_2\right\}}+\theta _1 n_{\left\{\tau _1,\tau
			_2,u_1,u_3\right\}}+n_{\left\{\tau _2,u_1\right\}} \\
		\left\{\tau _2,u_2\right\} & \theta _1 n_{\left\{\tau _1,u_3\right\}}+\theta _1
		n_{\left\{\tau _1,\tau _2,u_2,u_3\right\}}+n_{\left\{\tau _2,u_2\right\}} \\
		\left\{\tau _2,u_3\right\} & \theta _1 n_{\left\{\tau _1,u_2\right\}}+\theta _1
		n_{\left\{\tau _1,\tau _2,u_2,u_3\right\}}+n_{\left\{\tau _2,u_3\right\}} \\
		\left\{u_1,u_2\right\} & \theta _1 n_{\left\{\tau _1,u_3\right\}}+\theta _1
		n_{\left\{\tau _2,u_3\right\}}+\theta _1 n_{\left\{\tau _1,u_1,u_2,u_3\right\}}+\theta
		_1 n_{\left\{\tau _2,u_1,u_2,u_3\right\}}+n_{\left\{u_1,u_2\right\}} \\
		\left\{u_1,u_3\right\} & \theta _1 n_{\left\{\tau _1,u_2\right\}}+\theta _1
		n_{\left\{\tau _1,u_1,u_2,u_3\right\}}+n_{\left\{u_1,u_3\right\}} \\
		\left\{u_2,u_3\right\} & \theta _1 n_{\left\{\tau _2,u_1\right\}}+\theta _1
		n_{\left\{\tau _2,u_1,u_2,u_3\right\}}+n_{\left\{u_2,u_3\right\}} \\
		\left\{\tau _1,\tau _2,u_1,u_2\right\} & n_{\left\{\tau _1,\tau _2,u_1,u_2\right\}} \\
		\left\{\tau _1,\tau _2,u_1,u_3\right\} & n_{\left\{\tau _1,\tau _2,u_1,u_3\right\}} \\
		\left\{\tau _1,\tau _2,u_2,u_3\right\} & n_{\left\{\tau _1,\tau _2,u_2,u_3\right\}} \\
		\left\{\tau _1,u_1,u_2,u_3\right\} & n_{\left\{\tau _1,u_1,u_2,u_3\right\}} \\
		\left\{\tau _2,u_1,u_2,u_3\right\} & n_{\left\{\tau _2,u_1,u_2,u_3\right\}}
	\end{tabular}
	\caption{Chern number expansion for $ \theta = \theta_1 (( 
		0, 1),
		(  1,0 ),
		( -1,-1 )) $ (the 3-$\vec{q}$, triangular skyrmion lattice).}
	\label{tab:3q_2d}
	\end{table}
\end{turnpage}

\begin{turnpage}
	\begin{table}
		\scalebox{0.8}{
	\begin{tabular}{L|L}
		J' & \mathrm{Ch}_{J'} \\ \toprule
\{\} & \theta _1^3 n_{\left\{\tau _1,\tau _2,\tau _3,u_1,u_2,u_3\right\}}+\theta _1^2
n_{\left\{\tau _1,\tau _2,u_1,u_2\right\}}+\theta _1^2 n_{\left\{\tau _1,\tau
	_3,u_1,u_3\right\}}+\theta _1^2 n_{\left\{\tau _2,\tau _3,u_2,u_3\right\}}+\theta _1
n_{\left\{\tau _1,u_1\right\}}+\theta _1 n_{\left\{\tau _2,u_2\right\}}+\theta _1
n_{\left\{\tau _3,u_3\right\}}+n_{\{\}} \\
\left\{\tau _1,\tau _2\right\} & \theta _1 n_{\left\{\tau _3,u_3\right\}}+\theta _1
n_{\left\{\tau _1,\tau _2,\tau _3,u_3\right\}}+n_{\left\{\tau _1,\tau _2\right\}} \\
\left\{\tau _1,\tau _3\right\} & \theta _1 n_{\left\{\tau _2,u_2\right\}}+\theta _1
n_{\left\{\tau _1,\tau _2,\tau _3,u_2\right\}}+n_{\left\{\tau _1,\tau _3\right\}} \\
\left\{\tau _1,u_1\right\} & \theta _1^2 n_{\left\{\tau _2,\tau
	_3,u_2,u_3\right\}}+\theta _1^2 n_{\left\{\tau _1,\tau _2,\tau
	_3,u_1,u_2,u_3\right\}}+\theta _1 n_{\left\{\tau _2,u_2\right\}}+\theta _1
n_{\left\{\tau _3,u_3\right\}}+\theta _1 n_{\left\{\tau _1,\tau
	_2,u_1,u_2\right\}}+\theta _1 n_{\left\{\tau _1,\tau
	_3,u_1,u_3\right\}}+n_{\left\{\tau _1,u_1\right\}} \\
\left\{\tau _1,u_2\right\} & \theta _1 n_{\left\{\tau _3,u_3\right\}}+\theta _1
n_{\left\{\tau _1,\tau _3,u_2,u_3\right\}}+n_{\left\{\tau _1,u_2\right\}} \\
\left\{\tau _1,u_3\right\} & \theta _1 n_{\left\{\tau _2,u_2\right\}}+\theta _1
n_{\left\{\tau _1,\tau _2,u_2,u_3\right\}}+n_{\left\{\tau _1,u_3\right\}} \\
\left\{\tau _2,\tau _3\right\} & \theta _1 n_{\left\{\tau _1,u_1\right\}}+\theta _1
n_{\left\{\tau _1,\tau _2,\tau _3,u_1\right\}}+n_{\left\{\tau _2,\tau _3\right\}} \\
\left\{\tau _2,u_1\right\} & \theta _1 n_{\left\{\tau _3,u_3\right\}}+\theta _1
n_{\left\{\tau _2,\tau _3,u_1,u_3\right\}}+n_{\left\{\tau _2,u_1\right\}} \\
\left\{\tau _2,u_2\right\} & \theta _1^2 n_{\left\{\tau _1,\tau
	_3,u_1,u_3\right\}}+\theta _1^2 n_{\left\{\tau _1,\tau _2,\tau
	_3,u_1,u_2,u_3\right\}}+\theta _1 n_{\left\{\tau _1,u_1\right\}}+\theta _1
n_{\left\{\tau _3,u_3\right\}}+\theta _1 n_{\left\{\tau _1,\tau
	_2,u_1,u_2\right\}}+\theta _1 n_{\left\{\tau _2,\tau
	_3,u_2,u_3\right\}}+n_{\left\{\tau _2,u_2\right\}} \\
\left\{\tau _2,u_3\right\} & \theta _1 n_{\left\{\tau _1,u_1\right\}}+\theta _1
n_{\left\{\tau _1,\tau _2,u_1,u_3\right\}}+n_{\left\{\tau _2,u_3\right\}} \\
\left\{\tau _3,u_1\right\} & \theta _1 n_{\left\{\tau _2,u_2\right\}}+\theta _1
n_{\left\{\tau _2,\tau _3,u_1,u_2\right\}}+n_{\left\{\tau _3,u_1\right\}} \\
\left\{\tau _3,u_2\right\} & \theta _1 n_{\left\{\tau _1,u_1\right\}}+\theta _1
n_{\left\{\tau _1,\tau _3,u_1,u_2\right\}}+n_{\left\{\tau _3,u_2\right\}} \\
\left\{\tau _3,u_3\right\} & \theta _1^2 n_{\left\{\tau _1,\tau
	_2,u_1,u_2\right\}}+\theta _1^2 n_{\left\{\tau _1,\tau _2,\tau
	_3,u_1,u_2,u_3\right\}}+\theta _1 n_{\left\{\tau _1,u_1\right\}}+\theta _1
n_{\left\{\tau _2,u_2\right\}}+\theta _1 n_{\left\{\tau _1,\tau
	_3,u_1,u_3\right\}}+\theta _1 n_{\left\{\tau _2,\tau
	_3,u_2,u_3\right\}}+n_{\left\{\tau _3,u_3\right\}} \\
\left\{u_1,u_2\right\} & \theta _1 n_{\left\{\tau _3,u_3\right\}}+\theta _1
n_{\left\{\tau _3,u_1,u_2,u_3\right\}}+n_{\left\{u_1,u_2\right\}} \\
\left\{u_1,u_3\right\} & \theta _1 n_{\left\{\tau _2,u_2\right\}}+\theta _1
n_{\left\{\tau _2,u_1,u_2,u_3\right\}}+n_{\left\{u_1,u_3\right\}} \\
\left\{u_2,u_3\right\} & \theta _1 n_{\left\{\tau _1,u_1\right\}}+\theta _1
n_{\left\{\tau _1,u_1,u_2,u_3\right\}}+n_{\left\{u_2,u_3\right\}} \\
\left\{\tau _1,\tau _2,\tau _3,u_1\right\} & n_{\left\{\tau _1,\tau _2,\tau
	_3,u_1\right\}} \\
\left\{\tau _1,\tau _2,\tau _3,u_2\right\} & n_{\left\{\tau _1,\tau _2,\tau
	_3,u_2\right\}} \\
\left\{\tau _1,\tau _2,\tau _3,u_3\right\} & n_{\left\{\tau _1,\tau _2,\tau
	_3,u_3\right\}} \\
\left\{\tau _1,\tau _2,u_1,u_2\right\} & \theta _1 n_{\left\{\tau _3,u_3\right\}}+\theta
_1 n_{\left\{\tau _1,\tau _2,\tau _3,u_3\right\}}+\theta _1 n_{\left\{\tau _1,\tau
	_3,u_1,u_3\right\}}+\theta _1 n_{\left\{\tau _1,\tau _3,u_2,u_3\right\}}+\theta _1
n_{\left\{\tau _2,\tau _3,u_1,u_3\right\}}+\theta _1 n_{\left\{\tau _2,\tau
	_3,u_2,u_3\right\}}+\theta _1 n_{\left\{\tau _3,u_1,u_2,u_3\right\}}+\theta _1
n_{\left\{\tau _1,\tau _2,\tau _3,u_1,u_2,u_3\right\}}+n_{\left\{\tau _1,\tau
	_2,u_1,u_2\right\}} \\
\left\{\tau _1,\tau _2,u_1,u_3\right\} & n_{\left\{\tau _1,\tau _2,u_1,u_3\right\}} \\
\left\{\tau _1,\tau _2,u_2,u_3\right\} & n_{\left\{\tau _1,\tau _2,u_2,u_3\right\}} \\
\left\{\tau _1,\tau _3,u_1,u_2\right\} & n_{\left\{\tau _1,\tau _3,u_1,u_2\right\}} \\
\left\{\tau _1,\tau _3,u_1,u_3\right\} & \theta _1 n_{\left\{\tau _2,u_2\right\}}+\theta
_1 n_{\left\{\tau _1,\tau _2,\tau _3,u_2\right\}}+\theta _1 n_{\left\{\tau _1,\tau
	_2,u_1,u_2\right\}}+\theta _1 n_{\left\{\tau _1,\tau _2,u_2,u_3\right\}}+\theta _1
n_{\left\{\tau _2,\tau _3,u_1,u_2\right\}}+\theta _1 n_{\left\{\tau _2,\tau
	_3,u_2,u_3\right\}}+\theta _1 n_{\left\{\tau _2,u_1,u_2,u_3\right\}}+\theta _1
n_{\left\{\tau _1,\tau _2,\tau _3,u_1,u_2,u_3\right\}}+n_{\left\{\tau _1,\tau
	_3,u_1,u_3\right\}} \\
\left\{\tau _1,\tau _3,u_2,u_3\right\} & n_{\left\{\tau _1,\tau _3,u_2,u_3\right\}} \\
\left\{\tau _1,u_1,u_2,u_3\right\} & n_{\left\{\tau _1,u_1,u_2,u_3\right\}} \\
\left\{\tau _2,\tau _3,u_1,u_2\right\} & n_{\left\{\tau _2,\tau _3,u_1,u_2\right\}} \\
\left\{\tau _2,\tau _3,u_1,u_3\right\} & n_{\left\{\tau _2,\tau _3,u_1,u_3\right\}} \\
\left\{\tau _2,\tau _3,u_2,u_3\right\} & \theta _1 n_{\left\{\tau _1,u_1\right\}}+\theta
_1 n_{\left\{\tau _1,\tau _2,\tau _3,u_1\right\}}+\theta _1 n_{\left\{\tau _1,\tau
	_2,u_1,u_2\right\}}+\theta _1 n_{\left\{\tau _1,\tau _2,u_1,u_3\right\}}+\theta _1
n_{\left\{\tau _1,\tau _3,u_1,u_2\right\}}+\theta _1 n_{\left\{\tau _1,\tau
	_3,u_1,u_3\right\}}+\theta _1 n_{\left\{\tau _1,u_1,u_2,u_3\right\}}+\theta _1
n_{\left\{\tau _1,\tau _2,\tau _3,u_1,u_2,u_3\right\}}+n_{\left\{\tau _2,\tau
	_3,u_2,u_3\right\}} \\
\left\{\tau _2,u_1,u_2,u_3\right\} & n_{\left\{\tau _2,u_1,u_2,u_3\right\}} \\
\left\{\tau _3,u_1,u_2,u_3\right\} & n_{\left\{\tau _3,u_1,u_2,u_3\right\}} \\
\left\{\tau _1,\tau _2,\tau _3,u_1,u_2,u_3\right\} & n_{\left\{\tau _1,\tau _2,\tau
	_3,u_1,u_2,u_3\right\}} 
\end{tabular}}
\caption{Chern number expansion for the 3-$\vec{q}$ cubic hedgehog lattice in three dimensions with $\theta = \theta_1 \id_3$.}
\label{tab:3q_3d}
\end{table}
\end{turnpage}
\vfill

\newpage
\end{document}

%% file: commands.tex
\newcommand{\CM}{{\mathbb C}}

\newcommand{\NM}{{\mathbb N}}

\newcommand{\RM}{{\mathbb R}}

\newcommand{\TM}{{\mathbb T}}
\newcommand{\ZM}{{\mathbb Z}}

\newcommand{\Aa}{{\mathcal A}}

\newcommand{\Tt}{{\mathcal T}}

\newcommand{\Mm}{{\mathcal M}}

\newcommand{\Ss}{{\mathcal S}}

\newcommand{\deff}{ {d_\mathrm{eff}}}

\newcommand{\pgi}{Peter Gr\"unberg Institut and Institute for Advanced Simulation,
Forschungszentrum J\"ulich and JARA, 52425 J\"ulich, Germany}

\newcommand{\mainz}{Institute of Physics, Johannes Gutenberg University Mainz, 55099 Mainz, Germany}

\newcommand{\nyc}{
Department of Physics, Yeshiva University, New York, NY 10016, USA}

\renewcommand{\vec}[1]{\mathbf{#1}} 

\newcommand{\hatn}{\hat{\mathbf{n}}}

\newcommand{\abs}{\mathrm{\abs}}

\newcommand{\tr}{\mathrm{tr}}

\newcommand{\id}{\mathrm{id}}

\newcommand{\dd}{\mathrm{d}}

\newcommand{\bsigma}{\boldsymbol{\sigma}}

\newcommand{\xc}{\Delta_\mathrm{xc}}

